\documentclass[aps,pra,twocolumn,reprint,amsmath,amssymb]{revtex4-1}
\usepackage{graphicx}
\usepackage{dcolumn}
\usepackage{bm}
\usepackage[colorlinks,urlcolor=blue,citecolor=blue,linkcolor=blue,pdfstartview=FitH]{hyperref}
\usepackage{enumerate}
\usepackage{amsmath}
\usepackage{color}
\usepackage{cases}
\usepackage{microtype}
\usepackage{siunitx}
\usepackage{subdepth}

\usepackage[utf8]{inputenc}
\usepackage[T1]{fontenc}

\begin{document}

\title{Motion of vortices in inhomogeneous Bose--Einstein condensates}
\author{Andrew J. Groszek, David M. Paganin, Kristian Helmerson and Tapio P. Simula \\	
\textit{School of Physics and Astronomy, Monash University, Victoria 3800, Australia}}
		
\begin{abstract}
We derive a general and exact equation of motion for a quantised vortex in an inhomogeneous two-dimensional Bose--Einstein condensate. This equation expresses the velocity of a vortex as a sum of local ambient density and phase gradients in the vicinity of the vortex. We perform Gross--Pitaevskii simulations of single vortex dynamics in both harmonic and hard-walled disk-shaped traps, and find excellent agreement in both cases with our analytical prediction. The simulations reveal that, in a harmonic trap, the main contribution to the vortex velocity is an induced ambient phase gradient, a finding that contradicts the commonly quoted result that the local density gradient is the only relevant effect in this scenario. We use our analytical vortex velocity formula to derive a point-vortex model that accounts for both density and phase contributions to the vortex velocity, suitable for use in inhomogeneous condensates. Although good agreement is obtained between Gross--Pitaevskii and point-vortex simulations for specific few-vortex configurations, the effects of nonuniform condensate density are in general highly nontrivial, and are thus difficult to efficiently and accurately model using a simplified point-vortex description.
\end{abstract}

\maketitle

\section{Introduction}

Vortices are ubiquitous across a wide variety of physical contexts \cite{pismen_vortices_1999}, ranging from optical fields \cite{coullet_optical_1989, dennis_singular_2009} and free-electron waves \cite{uchida_generation_2010, verbeeck_production_2010, bliokh_theory_2017} to condensed matter systems such as superconductors \cite{blatter_vortices_1994} and superfluids \cite{vinen_detection_1961, donnelly_quantized_1991, bewley_superfluid_2006}. They arise in many interesting physical processes such as multi-wave interference \cite{nye_dislocations_1974}, phase transitions \cite{kibble_topology_1976, zurek_cosmological_1985, weiler_spontaneous_2008} and turbulence \cite{navon_emergence_2016}. As such, an understanding of their dynamics has applicability to a broad class of problems. Dilute gas Bose--Einstein condensates (BECs) present an ideal testbed for theoretically studying vortex physics, as the weak atomic interactions in these systems allow for a highly accurate mean-field description. In addition, there exist well established experimental techniques for creating \cite{matthews_vortices_1999, madison_vortex_2000, raman_vortex_2001, leanhardt_imprinting_2002, scherer_vortex_2007, wilson_experimental_2013, kwon_observation_2016} and imaging \cite{freilich_real-time_2010, wilson_situ_2015, seo_observation_2017} vortices in BECs, and hence laboratory studies of vortex physics in these systems are commonplace \cite{anderson_resource_2010}.

The simplest regime of vortex dynamics is that of a single vortex in a trapped BEC. An off-axis vortex has been experimentally observed to orbit the centre of a harmonically trapped condensate at a constant radius and frequency \cite{anderson_vortex_2000, bretin_quadrupole_2003, hodby_experimental_2003, freilich_real-time_2010, serafini_dynamics_2015}, and similar dynamics have been observed for vortices in superfluid Fermi gases \cite{yefsah_heavy_2013, ku_motion_2014}. Although conceptually simple, this motion has proved nontrivial to describe theoretically due to the inhomogeneous density profile which results from the harmonic trapping. Many attempts have been made to derive analytical expressions for the velocity of a single quantised vortex in these nonuniform systems \cite{jackson_vortex_1999, lundh_hydrodynamic_2000, svidzinsky_stability_2000, svidzinsky_dynamics_2000, fetter_vortex_2001, mcgee_rotational_2001, anglin_vortices_2002, sheehy_vortices_2004, al_khawaja_vortex_2005, nilsen_velocity_2006, jezek_vortex_2008, koens_perturbative_2012, dos_santos_hydrodynamics_2016, esposito_vortex_2017, biasi_exact_2017}; however, there is no consensus on the precise form of such an expression. In fact, even the specific physics responsible for the orbital motion is not universally agreed upon---there are conflicting descriptions of how density and phase gradients affect the vortex motion \cite{svidzinsky_stability_2000, sheehy_vortices_2004, nilsen_velocity_2006}, and there has been extensive debate over the relevance of image vortices to systems with soft boundaries \cite{anglin_vortices_2002, svidzinsky_stability_2000, mason_motion_2006, mason_motion_2008, jezek_vortex_2008, fetter_rotating_2009}. The effects of more general fluid inhomogeneity on vortex motion have also been studied theoretically \cite{mason_motion_2006, mason_motion_2008, cataldo_influence_2009, kevrekidis_vortex_2017}, a problem that will become increasingly relevant as experiments begin to utilise more complex trapping geometries \cite{henderson_experimental_2009, gaunt_bose-einstein_2013, navon_emergence_2016, gauthier_direct_2016}.

Despite the theoretical complications resulting from fluid inhomogeneity, focus has recently shifted towards increasingly complex regimes of vortex motion in effectively two-dimensional (2D) BECs. Experiments have been performed to investigate configurations such as vortex dipoles \cite{freilich_real-time_2010, neely_observation_2010, middelkamp_guiding-center_2011}, few-vortex clusters \cite{seman_three-vortex_2010, navarro_dynamics_2013}, and quantum turbulence \cite{neely_characteristics_2013, kwon_relaxation_2014, seo_observation_2017}. To theoretically model the dynamics of these 2D systems, it has proven fruitful to apply point-vortex approximations, in which the vortices are treated as point-particles whose motion is described by a set of coupled differential equations \cite{hess_angular_1967,  chang_dynamics_2002, middelkamp_bifurcations_2010, torres_dynamics_2011, torres_vortex_2011,  middelkamp_guiding-center_2011, navarro_dynamics_2013, simula_emergence_2014, murray_hamiltonian_2016, kim_role_2016}. These models, which are both conceptually and computationally simple, have been used to provide qualitative predictions of the dynamical and statistical behaviour observed in both experiments \cite{middelkamp_guiding-center_2011, navarro_dynamics_2013, moon_thermal_2015, kim_role_2016} and Gross--Pitaevskii simulations \cite{torres_vortex_2011, simula_emergence_2014, billam_spectral_2015, groszek_vortex_2018}. However, current point-vortex models cannot take general fluid inhomogeneity into account. In the case of harmonic trapping, a phenomenological term is commonly included to capture the vortex orbital motion (e.g.~\cite{middelkamp_guiding-center_2011, navarro_dynamics_2013}), but it only provides a quantitatively accurate prediction of the dynamics for vortices near the trap centre \cite{svidzinsky_stability_2000, fetter_rotating_2009}.
 
In this work, we use the Gross--Pitaevskii equation (GPE) to derive a general and exact expression for the velocity of a vortex, applicable in generic 2D Bose--Einstein condensates. Although this expression has appeared in previous BEC literature \cite{nilsen_velocity_2006, jezek_vortex_2008, dos_santos_hydrodynamics_2016} its importance has been understated. To demonstrate its accuracy and generality, we simulate the motion of a single vortex in both harmonic and hard-walled disk-shaped trapping potentials using the GPE. We find excellent agreement between the simulated dynamics and those predicted by the analytics. We also examine other models from the literature, and find that the expression derived here provides the best prediction of the vortex velocity. In addition, we show that it is possible to derive point-vortex equations of motion for arbitrary fluid geometries directly from this general equation, although approximations are necessary to account for ambient velocity fields that are induced by the inhomogeneous density.

This paper is structured as follows. In Sec.~\ref{sec:vortex_velocity}, we derive the vortex equation of motion, before verifying its accuracy using GPE simulations in Sec.~\ref{sec:numerical_study}. Section \ref{sec:single_vortex_literature} reviews past literature on the subject, and attempts to clarify a number of misconceptions present throughout previous works. In Sec.~\ref{sec:PVM_generalising}, we derive and test an improved point-vortex model for a harmonically trapped BEC. Finally, we summarise and discuss our findings in Sec.~\ref{sec:discussion}.

\section{The vortex velocity in an inhomogeneous superfluid \label{sec:vortex_velocity}}

The dynamical evolution of a Bose--Einstein condensate can be described using the nonlinear Schr\"{o}dinger equation $i \hbar \partial_t \psi= \mathcal{H} \psi$
with the Hamiltonian
\begin{equation} \label{eq:GPE_Hamiltonian}
\mathcal{H} = -\frac{\hbar^2}{2m} \nabla^2 + \mathcal{U}(\textbf{r},t),
\end{equation}
where $\psi$ is the condensate wavefunction, $m$ is the mass of the condensed atoms, and $\mathcal{U}$ is, in general, a complex operator. For the non-dissipative, zero temperature Gross--Pitaveskii model used throughout this work, $\mathcal{U}(\textbf{r},t) = V(\textbf{r},t) + g n(\textbf{r},t)$, where $V(\textbf{r},t)$ is an external trapping potential, $n(\textbf{r},t) \equiv |\psi(\textbf{r},t)|^2$ is the condensate density, and $g$ is a parameter that describes the interactions between condensate atoms. However, for the purposes of this derivation, the precise form of $\mathcal{U}$ turns out to be unimportant and could include terms due to thermal atom density or non-Hermitian growth and decay terms. Hence, the resulting equation for the vortex velocity is exceptionally general and its applicability is not limited to BECs.

We begin by assuming that at time $t=0$ there is a singly quantised vortex in a 2D condensate at the location $\textbf{r}_\circ=(x_\circ, y_\circ)$, which we express in complex notation as $z_\circ = x_\circ + i y_\circ$. Such a vortex state may be described, with no loss of generality, by the wavefunction
\begin{equation} \label{eq:total_wavefunction}
\psi_\circ \equiv \psi(\textbf{r},t=0)=(z- z_\circ) \tilde{\rho} \mathrm{e}^{i \tilde{\phi}},
\end{equation}
where $\tilde{\rho}(\textbf{r},t)$ and $\tilde{\phi}(\textbf{r},t)$ are smoothly varying real functions that, respectively, describe the background magnitude and phase of the wavefunction in the absence of the vortex. The function $z = x + iy$ accounts for both the density and phase of the condensate close to the vortex core.

We may use the Gross--Pitaevskii equation to propagate the wavefunction forward an infinitesimal time $\delta t$ by applying the unitary evolution operator:
\begin{subequations}
\begin{align} \label{eq:taylor_expansion}
\psi_{\rm new} \equiv \psi(\textbf{r},t=\delta t) &= \exp \left( -\frac{i}{\hbar} \mathcal{H} \delta t \right) \psi_\circ \\
&\approx \left( 1 -\frac{i}{\hbar} \mathcal{H} \delta t \right) \psi_\circ,
\end{align}
\end{subequations}
where in the second line we have expanded the exponential term in a Taylor series to first order in $\delta t$. Substituting the Hamiltonian, Eq.~\eqref{eq:GPE_Hamiltonian}, and the vortex ansatz wavefunction, Eq.~\eqref{eq:total_wavefunction}, into this expression results in
\begin{equation} \label{eq:taylor_expansion_2}
\psi_{\rm new} \approx (z-z_\circ) \tilde{\rho} \mathrm{e}^ {i \tilde{\phi}}
 - \frac{i}{\hbar} \delta t \left( -\frac{\hbar^2}{2m} \nabla^2 + \mathcal{U} \right) (z-z_\circ) \tilde{\rho} \mathrm{e}^ {i \tilde{\phi}}.
\end{equation}
The Laplacian term may be expanded to yield
\begin{align} \label{eq:laplacian}
\nabla^2 \big [ (z-z_\circ) \tilde{\rho} \mathrm{e}^ {i \tilde{\phi}} \big ] = & \bigg [ (z-z_\circ) \nabla^2 \tilde{\rho} + 2 (1, i) \cdot \nabla \tilde{\rho} \nonumber \\
&+ 2i \left \lbrace (z-z_\circ) \nabla \tilde{\rho} + \tilde{\rho} (1, i) \right \rbrace \cdot \nabla\tilde{\phi} \nonumber \\
&+ (z-z_\circ) \tilde{\rho} \left \lbrace i\nabla^2 \tilde{\phi} -(\nabla\tilde{\phi})^2 \right \rbrace \bigg ] \mathrm{e}^ {i \tilde{\phi}},
\end{align}
where we have used $\nabla (z - z_\circ) = (1, i)$ and $\nabla^2 (z - z_\circ) = 0$. Substituting Eq.~\eqref{eq:laplacian} into Eq.~\eqref{eq:taylor_expansion_2}, we evaluate $\psi_{\rm new}$ at $z = z_\circ + \delta z = (x_\circ + \delta x) + i (y_\circ + \delta y)$, which is the new location of the vortex after time $\delta t$. Because $\psi_{\rm new}$ must vanish at the new core location, we find that
\begin{align} \label{eq:expanded}
0 \approx \bigg \lbrace & \delta z\tilde{\rho} - \frac{i}{\hbar} \delta t  \bigg [ -\frac{\hbar^2}{2m} \bigg ( \delta z \nabla^2 \tilde{\rho} + 2i \left( \delta z \nabla \tilde{\rho} + \tilde{\rho} (1, i) \right) \cdot \nabla\tilde{\phi} \nonumber \\
&+ 2 (1, i) \cdot \nabla \tilde{\rho} + \delta z\tilde{\rho} \big (i\nabla^2 \tilde{\phi} -(\nabla\tilde{\phi})^2 \big) \bigg )  + \delta z \mathcal{U} \tilde{\rho} \bigg ] \bigg \rbrace \mathrm{e}^ {i \tilde{\phi}}.
\end{align}
The $\mathrm{e}^{i\tilde{\phi}}$ term is nonzero in general, and hence the term inside the braces must be equal to zero. We take the limit of the resulting expression as $\delta z \rightarrow 0$ and $\delta t \rightarrow 0$, leaving only terms that are first order in $\delta z$ and $\delta t$:
\begin{equation} \label{eq:dz_limit}
0 \approx \delta z\tilde{\rho} + \frac{i \hbar}{2m} \delta t \bigg ( 2i \tilde{\rho} (1, i) \cdot \nabla\tilde{\phi} + 2 (1, i) \cdot \nabla \tilde{\rho} \bigg ).
\end{equation}
Rearranging, we obtain an expression 
\begin{equation} \label{eq:velocity_1}
v_x + i v_y \equiv \frac{\delta z}{\delta t} = \frac{\hbar}{m} \bigg ( (1,i) \cdot \nabla\tilde{\phi} + (-i, 1) \cdot \frac{\nabla \tilde{\rho}}{\tilde{\rho}} \bigg ),
\end{equation}
for the vortex velocity $\bm{v}_v=(v_x, v_y)$ to first order accuracy, which becomes exact in the limit of adiabatic vortex motion \cite{simanek_adiabatic_1992, virtanen_adiabaticity_2001}. Expressed in vector form, the velocity of the vortex is
\begin{subequations} \label{eq:velocity_final}
\begin{align} 
\bm{v}_{v}(\textbf{r}_\circ) &= \frac{\hbar}{m} \left( \nabla \tilde{\phi} - \hat{\bm{\kappa}} \times \nabla \log \tilde{\rho} \right) \Big \vert_{\textbf{r}_\circ} \\
&\equiv \bm{v}_s(\textbf{r}_\circ) + \bm{v}_d(\textbf{r}_\circ).
\end{align}
\end{subequations}
Here we have identified two independent contributions to the vortex velocity: the background superfluid velocity due to ambient phase gradients $\bm{v}_s = (\hbar / m) \nabla \tilde{\phi}$, and a density gradient velocity $\bm{v}_d = -(\hbar / m) \hat{\bm{\kappa}} \times \nabla \log \tilde{\rho}$. In Eq.~(\ref{eq:velocity_final}a), we have explicitly included the dependence on the unit vector $\hat{\bm{\kappa}}$, which points in the direction of the vortex circulation vector $\bm{\kappa}=\kappa s \hat{\textbf{z}}$, where the integer $s$ is the vortex winding number, and $\kappa = h/m$ is the quantum of circulation. It is straightforward to verify this dependence on $\hat{\bm{\kappa}}$ by repeating the above calculation with $ z \rightarrow z^*$, $ z_\circ \rightarrow z_\circ^*$ and $\delta z \rightarrow \delta z^*$. We show in Sec.~\ref{sub:single_vortex_multi_quantum} that $\bm{v}_d$ is only dependent on the direction, and not the magnitude, of $\bm{\kappa}$.

We note that Eq.~\eqref{eq:velocity_final} is an entirely local expression---the vortex is not directly affected by global features of the condensate, such as its overall density profile, the presence of boundaries, or the existence of other vortices in the system. All such effects modify the motion of the vortex phase singularity implicitly through the changes in the ambient condensate density and phase. Furthermore, the vortex velocity derives exclusively from the kinetic energy term in the Hamiltonian, and hence the velocity of the vortex does not explicitly depend on $\mathcal{U}$ (although there is an implicit dependence via the wavefunction). Equation~\eqref{eq:velocity_final} is therefore generic and applies even for more general forms of $\mathcal{U}$, such as those which include dynamics of thermal atom densities, higher order nonlinear terms and dissipative effects.

\section{Numerical study of the velocity of a single vortex \label{sec:numerical_study}}

\subsection{The motion of a single vortex in an axisymmetric trap \label{sub:vortex_velocity_in_trap}}

The goal of Sec.~\ref{sec:numerical_study} is to verify the expression, Eq.~\eqref{eq:velocity_final}, for the vortex velocity by numerically simulating the motion of a single vortex in a trapped 2D BEC using the Gross--Pitaevskii equation. In doing so, we uncover a number of interesting features underlying the vortex motion, including the effects of varying density on the ambient superfluid velocity, and a multipole moment induced in the vortex velocity field. We consider two cylindrically symmetric geometries: a harmonic trap and a uniform disk-shaped trap with hard walls. It is well documented that, in each of these cases, a single off-centred vortex will orbit around the centre of the trap at a constant radius $r_\circ \equiv |\textbf{r}_\circ|$ with a radially dependent velocity $v_{\rm orb}(r_\circ)$ \cite{anderson_vortex_2000, svidzinsky_stability_2000,fetter_rotating_2009, freilich_real-time_2010}. However, this motion is typically thought to derive from different physical effects in each of these two cases.

In the uniform disk trap, the vortex motion is understood to arise from the Bernoulli effect, whereby the warping of the flow field due to the boundary leads to a pressure gradient, and hence a radial force, which drives the vortex in a circular path due to the gyroscopic effect of the rotating fluid. Equivalently, the motion can be described using the mathematical construction of image vortices---hypothetical vortex charges which exist outside the condensate and alter the fluid velocity field such that the boundary conditions of zero radial flow are satisfied \cite{viecelli_equilibrium_1995, fetter_rotating_2009}. These images generate a phase gradient within the fluid, and thus induce vortex motion via the first term in Eq.~\eqref{eq:velocity_final}.

By contrast, in the harmonic trap, the vortex orbital motion is usually attributed to the inhomogeneity of the condensate \cite{svidzinsky_stability_2000}, while the effect of the ambient superfluid velocity $\bm{v}_s$ has often been disregarded \cite{svidzinsky_stability_2000} or treated inadequately \cite{sheehy_vortices_2004, nilsen_velocity_2006} (see Secs.~\ref{sub:single_vortex_images} and \ref{sec:single_vortex_literature} for further discussion on previous results). However, our simulations reveal that both terms in Eq.~\eqref{eq:velocity_final} contribute significantly to the vortex velocity in the harmonic trap, as we show in Sec.~\ref{sub:results}.

\subsection{Numerical methods \label{sub:numerics}}

We numerically solve the Gross--Pitaevskii equation \cite{gross_structure_1961, pitaevskii_vortex_1961} using a fourth order split-step pseudospectral method on a $512 \times 512$ grid, with a spacing approximately equal to the healing length $\xi$. To obtain the harmonic and uniform disk geometries, we use trapping potentials $V_{h}(r) = \mu_h (r / R_h)^2$ and $V_{u}(r) = \mu_u (r / R_u)^{50}$, respectively, where the chemical potential in the harmonic trap is chosen to be times that in the uniform trap, $\mu_h =4 \mu_u$. We set the interaction parameter in the GPE to $g_h =2 g_u = 1.28\times10^4 \, \hbar^2 / m$, and use a trap radius of $R=128 \, \xi_h=64 \, \xi_u$, with $\xi_h = \xi_u / 2$. These parameter values ensure that we are well within the Thomas--Fermi regime, and physically, could for example correspond to a $^{87}\mathrm{Rb}$ BEC in a trap with 2D radius $R = \SI{30}{\micro\meter}$. An axial radius of $R_z = 0.1 R = \SI{3}{\micro \meter}$ in each trap would then correspond to a total atom number of $N_h = 2 N_u \approx 1.3 \times 10^6$, assuming harmonic confinement in the $z$-direction. For each trap, we calculate the ground state using imaginary time propagation. We then imprint a vortex of charge $s$ at location $\textbf{r}_\circ$ by multiplying the wavefunction by $f(|\textbf{r} - \textbf{r}_\circ|)\mathrm{e}^{i \phi_v(\textbf{r})}$, where $\phi_v(\textbf{r}) = s\arctan \left[ (y - y_\circ)/(x - x_\circ) \right]$, and $f(x) = x/\sqrt{x^2 + 2\xi^2}$ is the approximate density profile of a vortex \cite{pethick_bose-einstein_2008}. This initial state is evolved to $t=5 \times 10^4 \, \hbar/\mu$ using the GPE (long enough to see at least four orbits at the lowest frequencies). As a result of the imprinting method, the ambient phase $\tilde{\phi}(\textbf{r})$ is initially zero everywhere. When the initial state is evolved in time, the ambient phase field develops continuously over a small fraction ($\lesssim 1\%$) of a vortex orbital period. During this time, the vortex accelerates from rest until it reaches its (approximately) constant angular frequency and radius. Vortices are identified by locating phase singularities in the wavefunction.

Throughout the time evolution, we independently measure each of the three terms in Eq.~\eqref{eq:velocity_final}:
\begin{enumerate}[(i)]
\item The total orbital velocity [the left-hand side of Eq.~\eqref{eq:velocity_final}] is calculated from the angular frequency of the vortex orbital motion as $v_{\rm orb} = \omega_{\rm orb} r_\circ$.

\item To measure the ambient superfluid velocity field $\bm{v}_s(\textbf{r}_\circ)= (\hbar/m) \nabla \tilde{\phi}(\textbf{r}) \vert _{\textbf{r}_\circ}$, we first calculate the ambient phase $\tilde{\phi}(\textbf{r})$ by subtracting the axisymmetric vortex phase field from the total phase of the condensate: $\tilde{\phi}(\textbf{r}) = \phi(\textbf{r}) - \phi_v(\textbf{r})$. This subtraction must be done carefully to minimise numerical fluctuations at the vortex core. We then average the resulting velocity field $\bm{v}_s(\textbf{r})$ within a series of annuli $r_a - \xi < |\textbf{r} - \textbf{r}_\circ| < r_a + \xi$ around the vortex core, where $r_a$ is varied between $2 \, \xi$ and $11 \, \xi$. Due to fluctuations in the velocity within $|\textbf{r} - \textbf{r}_\circ| \lesssim \xi$ (and contributions from a multipole velocity field---see Sec.~\ref{sub:multipole}), we extrapolate the measurements from the larger annuli to determine the velocity at $\textbf{r}_\circ$.

\item The density-dependent velocity $\bm{v}_d (\textbf{r}_\circ) = - (\hbar/m) \hat{\bm{\kappa}} \times \nabla \tilde{\rho} / \tilde{\rho} \vert _{\textbf{r}_\circ}$ is measured numerically around the vortex core by fitting a plane $P(x,y) = A + Bx + Cy$ to $\rho(\textbf{r})=|\psi(\textbf{r})|$ within the annuli $r_a - \xi < |\textbf{r}-\textbf{r}_\circ| < r_a + \xi$, where $r_a$ is varied between $6 \, \xi$ and $11 \, \xi$. We then calculate the density terms as: $\tilde{\rho}(\textbf{r}_\circ) = \langle A \rangle$, $|\nabla \tilde{\rho}|_{\textbf{r}_\circ} =  ( \langle B \rangle^2 + \langle C \rangle^2 )^{1/2}$, where the average is taken over both time and the radii $r_a$. For comparison, we also calculate $\bm{v}_d$ using the ground state density profile, and find very good agreement between the two methods.
\end{enumerate}

\subsection{Results \label{sub:results}}
\subsubsection{Vortex orbital dynamics \label{sub:single_vortex_results}}
\begin{figure}[tb]
\centering
\includegraphics[width=\columnwidth]{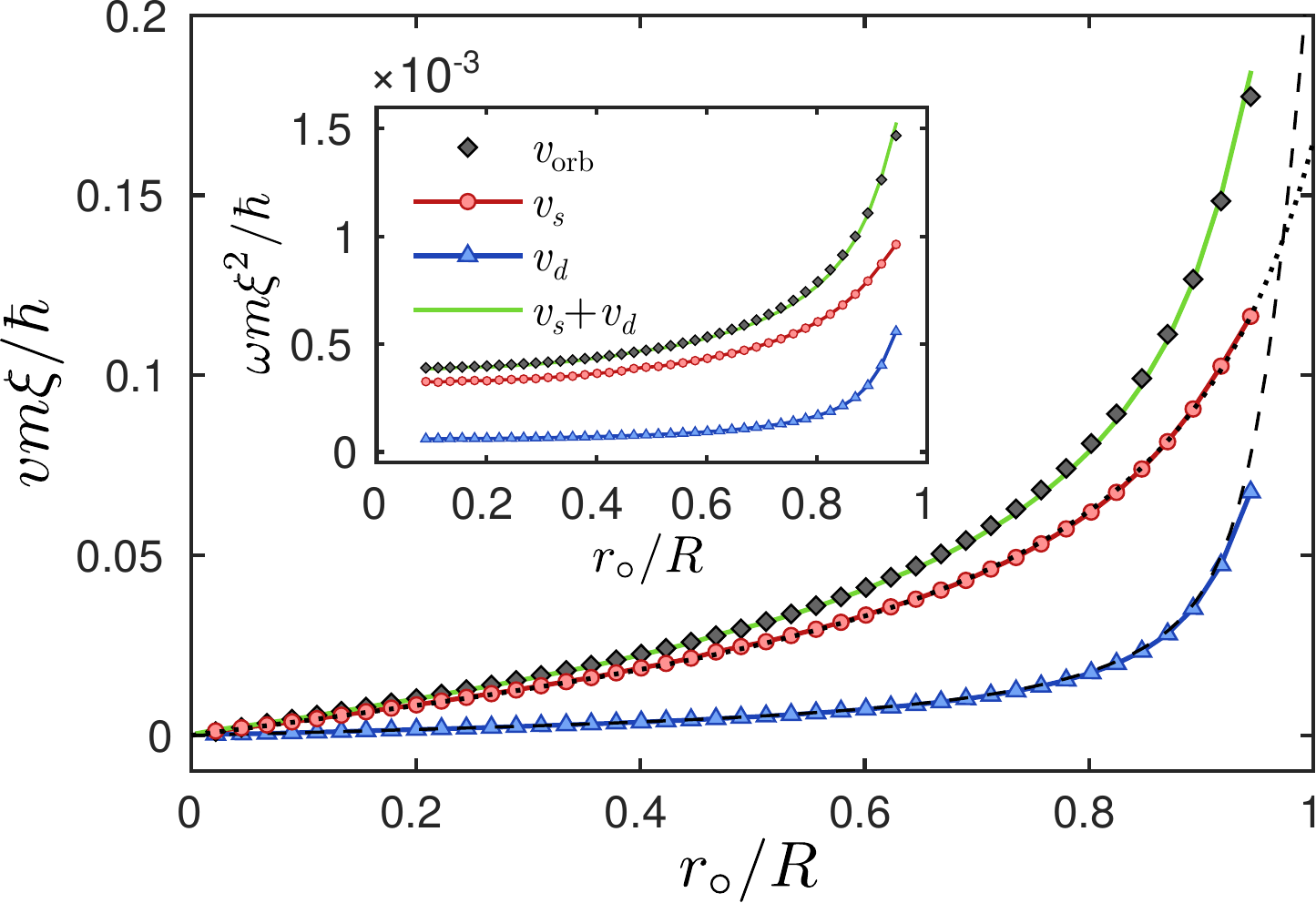}
\caption{The contributions to the vortex orbital velocity and frequency (inset) in a harmonically trapped condensate, for a vortex initiated at variable radius $r_\circ$. In the main frame the black diamonds denote the measured orbital velocity $v_{\rm orb}$, while the two terms on the right hand side of Eq.~\eqref{eq:velocity_final}, $v_s$ and $v_d$, are plotted as red circles and blue triangles, respectively. The sum $v_s + v_d$ is also shown as a solid green line for comparison with $v_{\rm orb}$. All corresponding frequencies are plotted equivalently in the inset. In the main frame, the dotted line shows the fit $v(r_\circ) =(\hbar / m) \alpha r_\circ / (\beta R^2 - r_\circ^2)$, a generalised image vortex velocity, to $v_s(r_\circ)$, where $\alpha=6.79$, $\beta=1.32$ (see Sec.~\ref{sub:PVM_derivation}). The dashed curve is the result of calculating $v_d$ using the ground state density profile. In the inset, the data for the lowest four radii have been omitted due to numerical noise.}
\label{fig:figure_1_single}
\end{figure}

The numerically measured velocity curves for a vortex located at variable radius $r_\circ$ in a harmonically trapped system are shown in Fig.~\ref{fig:figure_1_single}. As predicted by Eq.~\eqref{eq:velocity_final}, the sum of the density and phase gradient terms gives excellent agreement with the total vortex velocity. For improved clarity at small values of $r_\circ$, we have also included the orbital frequency measurements in the inset of the Figure. This data clearly shows that, for all radii, the ambient superfluid velocity is actually the dominant contribution to the vortex motion, while the density-dependent effect only becomes significant near the boundary. This finding is in contradiction with much of the literature on the topic, as we discuss in Sec.~\ref{sec:single_vortex_literature}.

\begin{figure}[tb]
\centering
\includegraphics[width=\columnwidth]{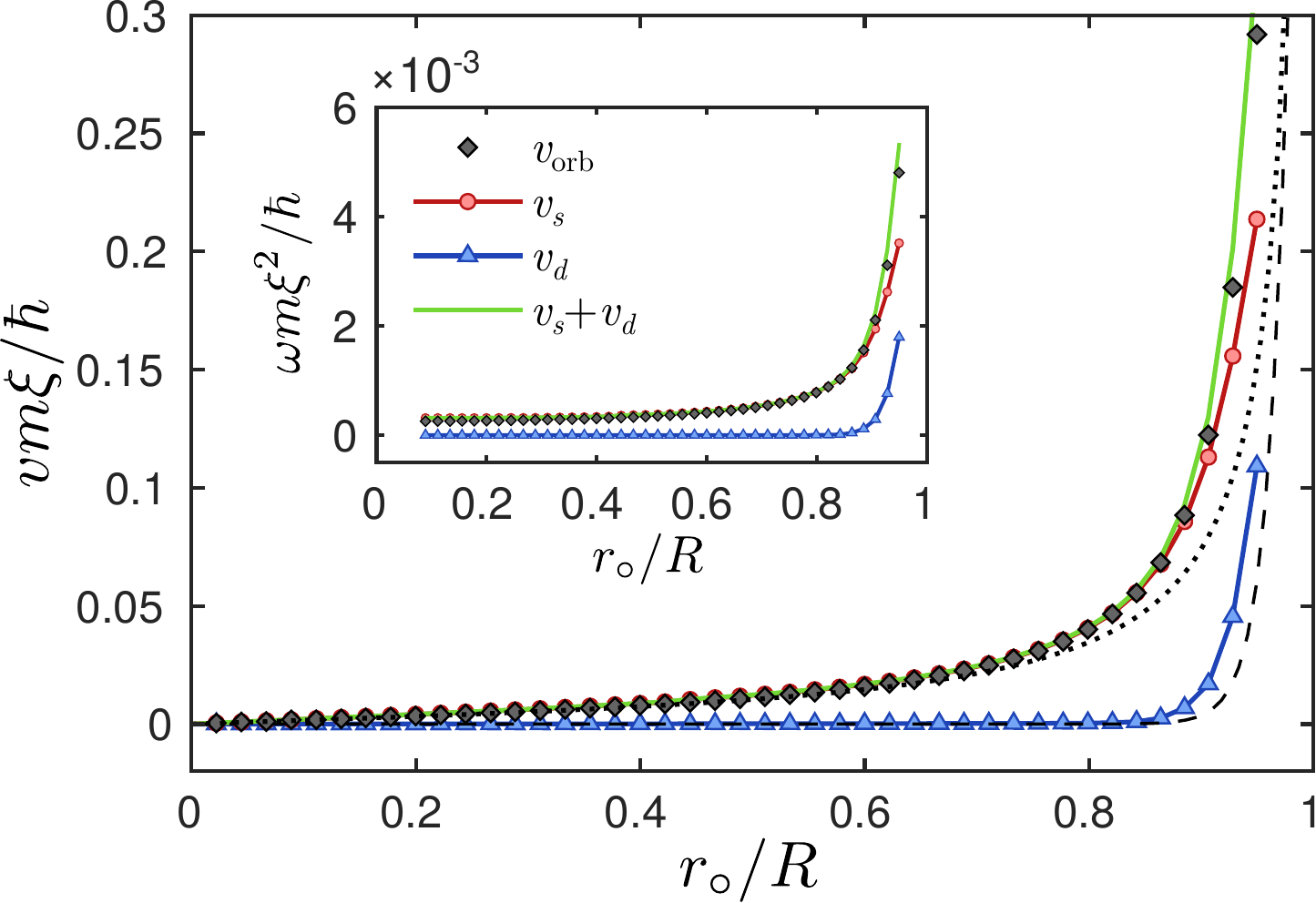}
\caption{The contributions to the vortex orbital velocity and frequency (inset) in a uniform, disk-shaped condensate, for a vortex imprinted at variable radius $r_\circ$. The data are labelled as in Fig.~\ref{fig:figure_1_single}, except that the dotted curve shown here 
is the velocity $v(r_\circ) = (\hbar / m) r_\circ / (R^2 - r_\circ^2)$ produced by an image vortex at $\bar{r}_\circ = R^2/r_\circ$. As in Fig.~\ref{fig:figure_1_single}, the frequency data at the lowest four radii have been omitted due to numerical noise.}
\label{fig:figure_2_single_uniform}
\end{figure}

Figure \ref{fig:figure_2_single_uniform} shows the measured velocity data for a single vortex in the uniform trap. Once again, we find that the total velocity is well described by the sum of the phase and density terms, as Eq.~\eqref{eq:velocity_final} predicts. We also observe that, in this system, the overwhelming contribution to the vortex velocity for radii $r_\circ \lesssim 0.9 \, R$ is the phase gradient. This is to be expected, since a vortex should move with the background flow field in a uniform superfluid~\cite{fetter_vortices_1966}. The sudden increase in $v_d$ near the boundary is due to the finite width of the wall---in an infinite cylindrical well, this term would remain negligible everywhere. We also find that, for small radii, $v_s(r_\circ)$ agrees well with the velocity field produced by an image vortex outside the condensate at radius $\bar{\textbf{r}}_\circ=\textbf{r}_\circ R^2/ |\textbf{r}_\circ|^2$, the expected image location for a disk-shaped system with infinitely hard walls \cite{pointin_statistical_1976, viecelli_equilibrium_1995}. As the vortex approaches the edge of the fluid, the phase gradient velocity becomes stronger than the image vortex predicts. This can be attributed to the fact that neither the vortex nor the wall are infinitesimally narrow features and consequently the ideal point-vortex image picture fails near the boundary of the condensate.

\subsubsection{Contributions to the ambient velocity field \label{sub:single_vortex_images}}

\begin{figure}[bt]
\centering
\includegraphics[width=\columnwidth]{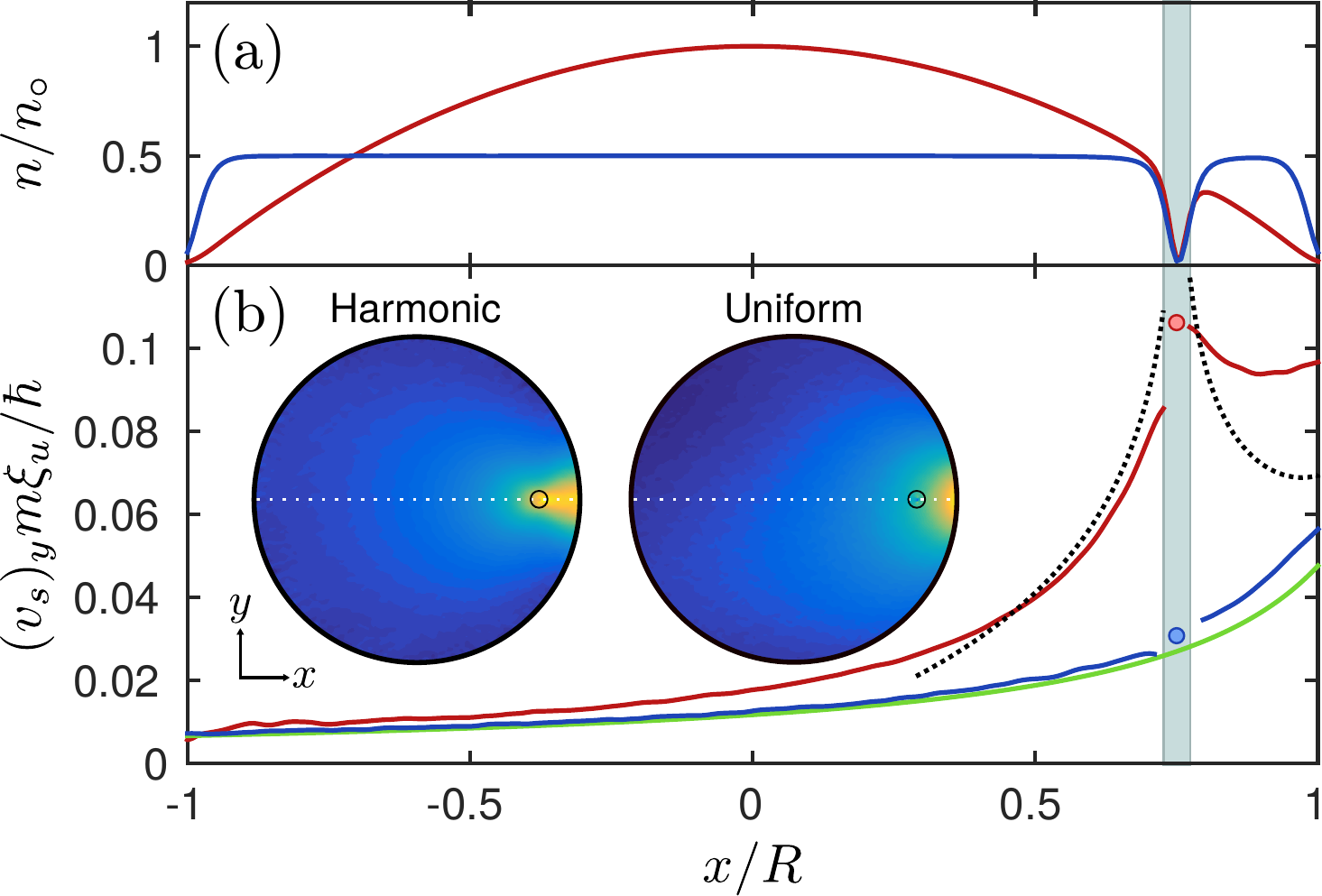}
\caption{The (a) one-dimensional density profile $n(x)=|\psi(x,0)|^2$ and (b) $y$-component of the background velocity field $\bm{v}_s=\nabla \tilde{\phi}$ along the $x$-axis in both the harmonic (red) and uniform disk (blue) traps for a vortex at position $x_\circ \approx 0.75 \, R$ (highlighted by the vertical shaded region). In the inset of panel (b), the $y$-component of $\bm{v}_s$ has been plotted across the whole condensate for each trap, with a dotted line indicating the cross-section shown in the main frame, and a black circle denoting the vortex location. The colour scales in the inset are the same as the $y$-axis of (b). All numerical data has been averaged over $\sim 130$ dynamical frames in each geometry. The solid green line in (b) is the velocity field produced by an image vortex at $\bar{x}_\circ \approx 0.75^{-1} \, R$, while the black dotted line shows the sum of Eq.~\eqref{eq:velocity_sheehy} and the image vortex velocity field. For comparison with Figs.~\ref{fig:figure_1_single} and \ref{fig:figure_2_single_uniform}, the measurements of $v_s$ at $r_\circ \approx 0.75 \, R$ in each trap are also shown as filled circles (note that there is a factor of two difference for the velocity in the harmonic trap due to the scaling with $\xi_u$).}
\label{fig:figure_3_vfields}
\end{figure}

Whereas the density gradient velocity in Eq.~\eqref{eq:velocity_final} is straightforward to measure from ground state properties, the ambient velocity field $\bm{v}_s(\textbf{r})$ induced by the vortex is, in general, more complicated. To demonstrate this, we measure the background velocity field everywhere in the condensate for a vortex at radius $r_\circ \approx 0.75 \, R$ in each of our two traps. The inset of Fig.~\ref{fig:figure_3_vfields}(b) shows the $y$-component of each measured velocity field over the entire condensate when the vortex is located at $\textbf{r}_\circ \approx (0.75 \, R, 0)$, while the main frame of panel (b) shows a one-dimensional slice through this field along the $x$-axis. Panel (a) shows the corresponding density profiles, normalised to $n_\circ$, the maximum density in the harmonic trap.

In the uniform trap, the background velocity field is well described by an image vortex located at $\bar{\textbf{r}}_\circ \approx (0.75^{-1} \, R, 0)$ (the expected location for a hard-walled disk trap), although the agreement becomes worse near the boundary closest to the vortex, due to the finite core size and boundary width. By contrast, the velocity field in the harmonic trap is more complicated. A peak in the background velocity in the region around the vortex core is clearly visible, and has been previously identified and discussed in Ref.~\cite{jezek_vortex_2008}. It was suggested in Ref.~\cite{jezek_vortex_2008} that the background velocity field $\bm{v}_s(\textbf{r})$ could be split into two independent contributions: an image vortex field arising from the presence of the boundary, plus an additional contribution due to the fluid inhomogeneity at the vortex location. In fact, Sheehy and Radzihovsky \cite{sheehy_vortices_2004} derived an approximate expression for this second contribution,
\begin{equation} \label{eq:velocity_sheehy}
\bm{v}_{\textrm{peak}}(\textbf{r}) = \frac{\hbar}{m} \hat{\textbf{z}} \times \frac{\nabla \tilde{\rho}(\textbf{r}_\circ)}{\tilde{\rho}(\textbf{r}_\circ)} \log \left( \frac{|\textbf{r}-\textbf{r}_\circ | |\nabla \tilde{\rho}(\textbf{r}_\circ)|}{|\tilde{\rho}(\textbf{r}_\circ)|} \right),
\end{equation}
which is responsible for the peak in the region around the vortex \footnote{It was assumed in their derivation that this was the only contribution to the vortex orbital velocity, which we have shown is not the case}. For comparison, we show in Fig.~\ref{fig:figure_3_vfields}(b) the sum of the image velocity field and Eq.~\eqref{eq:velocity_sheehy}, as suggested in Ref.~\cite{jezek_vortex_2008}. While qualitatively reasonable, this approach does not provide quantitative accuracy. Moreover, Eq.~\eqref{eq:velocity_sheehy} is only valid near, but outside of, the core region, and therefore fails at greater distances.

Interpreting these observations in light of Eq.~\eqref{eq:velocity_final}, we emphasise that a density gradient at the vortex location produces two distinct effects on the vortex motion:
\begin{enumerate}[(i)]
\item A `direct' effect on the vortex produced by $\bm{v}_d$ [which does not contribute to the ambient velocity field $\bm{v}_s$ shown in Fig.~\ref{fig:figure_3_vfields}(b)].
\item An `indirect' effect via a warping of the phase field which enters $\bm{v}_s$ in addition to an image effect due to the boundary, and which manifests as a peak in the azimuthal velocity field around the vortex in the harmonically trapped condensate [shown in Fig.~\ref{fig:figure_3_vfields}(b)].
\end{enumerate}

Unlike for the uniform trap, we do not expect the background `image vortex' field in an inhomogeneous system to be described by a single image point-vortex located outside the fluid. Instead, we expect the softness of the boundary to delocalise the image, much like a spherical aberration produced by a soft mirror \cite{rose_geometrical_2009}. It may therefore be possible to approximate the image field more accurately using a configuration of multiple image vortices; however, doing so would destroy the simplified physical picture that makes the image representation appealing.

\subsubsection{Induced multipole moments \label{sub:multipole}}

In addition to the effects of boundaries and varying condensate density on the background velocity field $\bm{v}_s(\textbf{r})$ (discussed in Sec.~\ref{sub:single_vortex_images}), 
dipole, and higher multipole, moments in the velocity field $\bm{v}_i(\textbf{r})$ of the vortex have been predicted to emerge as a result of the internal structure of the defect. This effect arises due to the dynamical excitation of the $n_z=0$ kelvon quasiparticles localised within the vortex core \cite{pitaevskii_vortex_1961, dodd_excitation_1997, isoshima_bose-einstein_1997, virtanen_structure_2001, fetter_kelvin_2004, simula_kelvin_2008}. Because the vortices considered here are two-dimensional, kelvons with axial quantum numbers $n_z>0$ are suppressed~\cite{rooney_suppression_2011}.

In Ref.~\cite{klein_internal_2014}, it was predicted that a vortex moving relative to the background superflow should exhibit an altered intrinsic velocity field $\bm{v}_i(\textbf{r})$ which is no longer circularly symmetric. Outside of the vortex core, the corrections can be expressed in terms of a multipole expansion \cite{klein_internal_2014}:
\begin{align} \label{eq:dipole_expansion}
\bm{v}_i(\textbf{r}) = & \bm{v}_i^{(1)}(\textbf{r}) + \bm{v}_i^{(2)}(\textbf{r}) + \ldots \nonumber \\
 = & \frac{\hbar}{m} \bigg [ \hat{\textbf{z}} \times \frac{\textbf{r} - \textbf{r}_\circ}{|\textbf{r} - \textbf{r}_\circ|^2} \nonumber \\
 & +  \frac{(\textbf{r} - \textbf{r}_\circ)^2 \textbf{d} - 2 \left[ \textbf{d} \cdot (\textbf{r} - \textbf{r}_\circ) \right] (\textbf{r} - \textbf{r}_\circ) }{|\textbf{r} - \textbf{r}_\circ|^4} + \ldots \bigg ],
\end{align}
where the dipole moment
\begin{equation} \label{eq:dipole_moment}
\textbf{d} \equiv \bm{v}_{\rm rel} \frac{m \xi^2}{\hbar} \left( \log \left | \frac{\textbf{r}-\textbf{r}_\circ}{\xi} \right| - a \log \left| \frac{m \xi \bm{v}_{\rm rel}}{\hbar} \right| \right).
\end{equation}
Here, $a \approx 1.49$ is a numerical constant, and $\bm{v}_{\rm rel}$ is the velocity of the vortex relative to the superfluid in the vortex frame of reference.

\begin{figure}[t]
\centering
\includegraphics[width=\columnwidth]{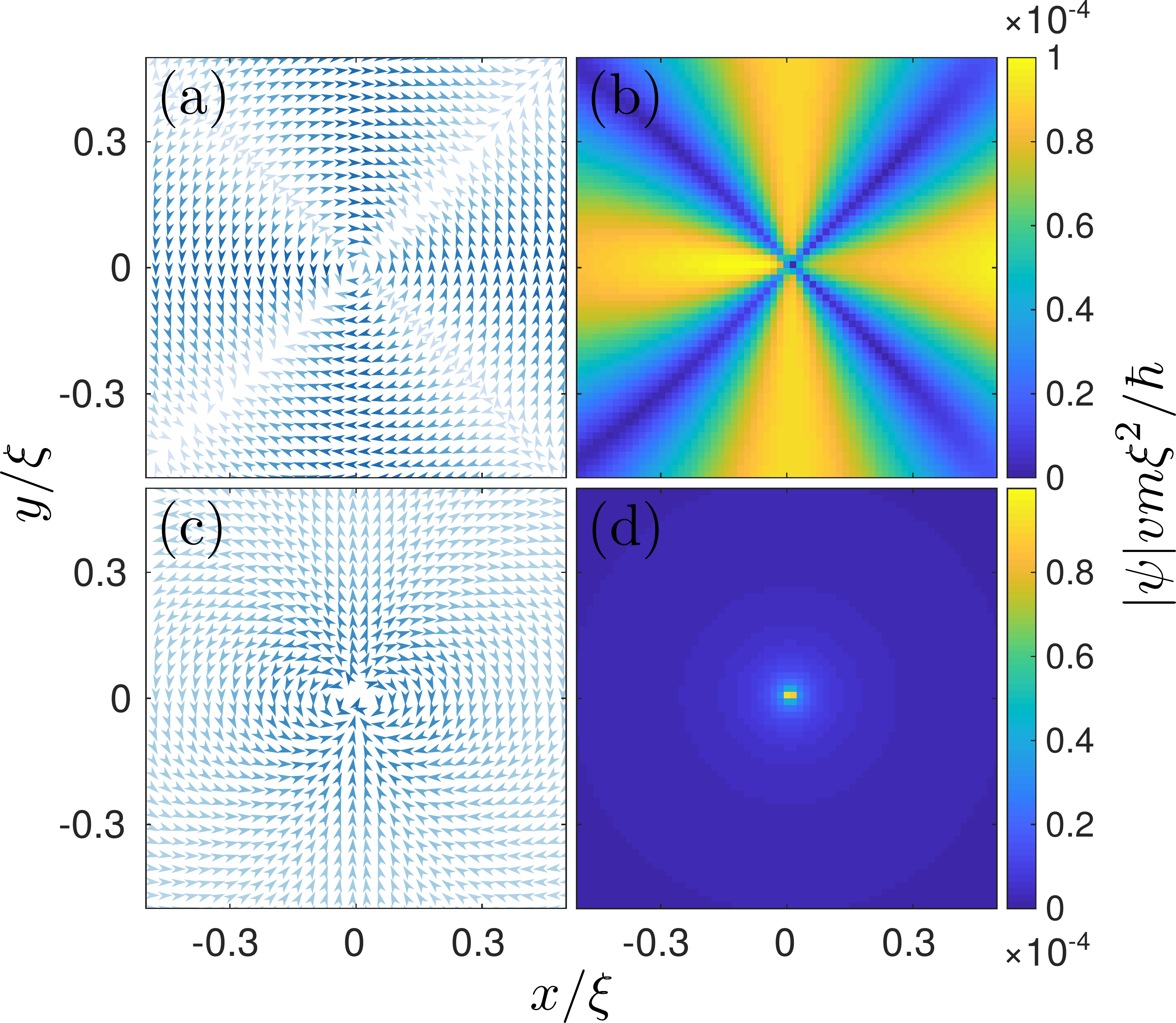}
\caption{Comparison between the numerical [(a)/(b)] and predicted [(c)/(d)] density-weighted velocity fields within the vortex core, left over after subtracting out the vortex monopole field $\bm{v}_i^{(1)}(\textbf{r})$ and the local background velocity $\left \langle \bm{v}_s \right \rangle$ (averaged over the region shown). The left and right columns, respectively, show the direction and magnitude of each velocity field. The vortex is located at $x_\circ \approx 0.5 \, R$, and will travel in the positive $y$-direction under real time evolution.}
\label{fig:figure_3_multipole}
\end{figure}

To investigate the possibility of such multipole effects in our Gross--Pitaevskii simulations, we have performed further numerical calculations in the disk-shaped trap, using an increased resolution of $4096 \times 4096$ grid points, and a smaller interaction parameter, $g = 148 \, \hbar^2 / m$. This reduces the condensate radius to $R \approx 21 \, \xi$, and increases the number of grid points per healing length to $\sim 64$. After imprinting the vortex phase winding into the ground state of the trap and evolving for a short amount of imaginary time, a quadrupole-like structure becomes visible in the flow field, once both the monopole field $\bm{v}_i^{(1)}(\textbf{r})$ and the local mean background velocity $\left \langle \bm{v}_s \right \rangle$ have been subtracted away \footnote{Strictly, the induced multipole moments are intrinsic to the vortex `particle' and could therefore be removed from the phase field before calculating the smooth background field $\bm{v}_s = \nabla \tilde{\phi}$ which drives the vortex motion. However, since we have only subtracted the circularly symmetric monopole component $\bm{v}_i^{(1)}(\textbf{r})$, the higher order multipole contributions remain in our measured `background' field $\bm{v}_s(\textbf{r})$.}. Figure~\ref{fig:figure_3_multipole}(a)--(b) shows this numerically measured velocity field for a vortex initiated at $\textbf{r}_\circ \approx (0.5 \, R, 0)$. Although the data shown has been obtained using imaginary time propagation, the same structure develops during real time evolution, and is 1-2 orders of magnitude weaker than the background superflow $\bm{v}_s$ driving the vortex motion.

We are only able to reproduce a dipole field---such as the prediction of Eqs.~\eqref{eq:dipole_expansion} and \eqref{eq:dipole_moment} shown in Fig. \ref{fig:figure_3_multipole}(c)--(d)---as a numerical artifact arising from an inaccurate subtraction of the monopole field, which essentially imprints a vortex--antivortex dipole in the wavefunction. Further investigation into the vortex core localised multipolar velocity fields is a topic of future work.

\subsubsection{The velocity of a vortex with multiple circulation quanta \label{sub:single_vortex_multi_quantum}}

\begin{figure}[t]
\centering
\includegraphics[width=\columnwidth]{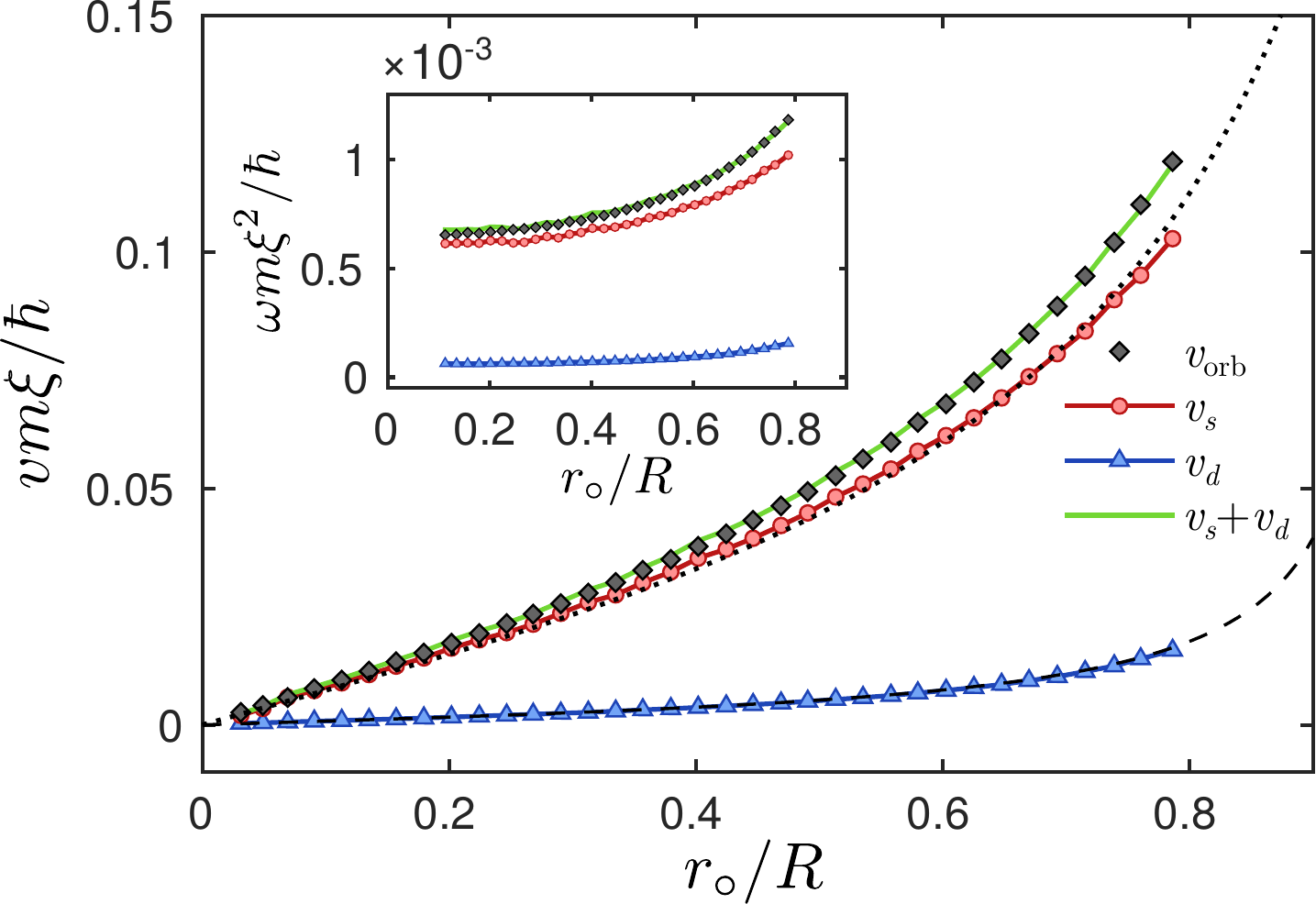}
\caption{The contributions to the orbital velocity and frequency (inset) for a charge $s=2$ vortex in a harmonic trap, initiated at variable radius $r_\circ$. The data are labelled as in Figs.~\ref{fig:figure_1_single} and \ref{fig:figure_2_single_uniform}. The dotted curve shown in the main frame is a fit to $v(r_\circ) = (\hbar / m) \alpha r_\circ / (1.32 R^2 - r_\circ^2)$, which gives $\alpha=12.26$, a value that is $\sim 1.81$ times larger than that obtained from the single vortex fit. Frequency data at the lowest radii have been omitted due to numerical fluctuations.}
\label{fig:figure_3_multiquantum}
\end{figure}

To confirm that Eq.~\eqref{eq:velocity_final} applies equally well for higher charge vortices, we have repeated our numerical analysis of the vortex velocity in a harmonic trap using a single $s=2$ vortex. Due to the inherent energetic instabilities of this vortex state~\cite{rokhsar_vortex_1997, leanhardt_imprinting_2002}, the singularity immediately splits into two singly-charged vortices, which  continuously emit phonons and gradually drift apart, causing the centre-of-mass velocity to decrease (for approximately one trap orbit, however, the two vortex cores are indiscernible). To minimise the effects of this splitting on our velocity data, we cut off our measurements once the distance $d_{v}$ between the two singularities becomes greater than $8 \, \xi$, and only calculate the background fields for the early times when $d_{v} \leq 3 \, \xi$. The obtained velocity and frequency curves are shown in Fig.~\ref{fig:figure_3_multiquantum}, demonstrating that Eq.~\eqref{eq:velocity_final} still holds, even for a multi-quantum vortex. Surprisingly, if the derivation in Sec.~\ref{sec:vortex_velocity} is repeated using an ansatz wavefunction with $(z-z_\circ) \rightarrow (z-z_\circ)^{|s|}$ (i.e.~a multi-quantum vortex of charge $s>0$), then the velocity in Eq.~\eqref{eq:velocity_final} becomes $\bm{v}_v \rightarrow |s| (\bm{v}_s + \bm{v}_d)$, which does not match with our numerical results.

For all radii, the total orbital velocity of the vortex is approximately $1.6$ times greater than the velocity obtained for a charge $s=1$ vortex at the same radius. This increase comes entirely from the phase gradient term, which grows by $\approx 1.8$ times---slightly lower than the factor of two one would expect from a simple image vortex picture. We have confirmed that, in the uniform disk trap, the $v_s$ component does scale by a factor of two, suggesting that the slightly smaller value observed in the harmonic trap is related to the shape of the induced velocity peak discussed in Sec.~\ref{sub:single_vortex_images}. It is interesting to note that, for vortices with large circulation, the phase gradient term in Eq.~\eqref{eq:velocity_final} becomes increasingly dominant, since $\bm{v}_d$ does not scale with $|s|$.

\section{Comparison with results in the literature \label{sec:single_vortex_literature}}

Many expressions describing the motion of vortices in inhomogeneous fluids to varying degree of accuracy are found in the literature. We find that, unlike our analytical solution Eq.~\eqref{eq:velocity_final}, none of the other models agree precisely with the numerically measured orbital velocity of a single vortex. In the following, we discuss the two most widely used approaches, and briefly review some more recent results.

\subsection{The two standard approaches} \label{sub:literature_models}
The first of the two common methods from the literature invokes a force balancing argument whereby the negative gradient of the energy $E(\textbf{r}_\circ)$ is equated to the `Magnus force' on the vortex \cite{jackson_vortex_1999, mcgee_rotational_2001, sheehy_vortices_2004, ku_motion_2014, toikka_asymptotically_2017, guenther_quantized_2017}:
\begin{equation} \label{eq:velocity_literature_energy}
\textbf{F}_{\rm Mag} \stackrel{?}{=} m \tilde{n} \bm{\kappa} \times  \bm{v}_v = \nabla E ( \textbf{r}_\circ),
\end{equation}
where $\tilde{n} \equiv \tilde{\rho}^2$, and the gradient $\nabla E ( \textbf{r}_\circ)$ is taken with respect to the vortex location $\textbf{r}_{\circ}$. The same formula has also been obtained using a variational Lagrangian approach \cite{svidzinsky_stability_2000, lundh_hydrodynamic_2000}. The advantage of this expression is that the vortex velocity can be calculated directly from the total energy $E$ of the fluid, which is straightforward to measure numerically, and can be approximated analytically for a single vortex \cite{jackson_vortex_1999, svidzinsky_stability_2000, lundh_hydrodynamic_2000, mcgee_rotational_2001}. However, we argue that this approach also has a number of significant shortcomings. Firstly, Eq.~\eqref{eq:velocity_literature_energy} requires knowledge of the global properties of the condensate, making it less general than the local description of Eq.~\eqref{eq:velocity_final}. Moreover, as suggested by the $\stackrel{?}{=}$ notation, the Magnus force, rather than being proportional to the vortex velocity, should be proportional to the velocity of the vortex \textit{relative} to the background superflow \cite{ambegaokar_dynamics_1980, ao_berrys_1993, thouless_transverse_1996}:
\begin{equation} \label{eq:magnus_force}
\textbf{F}_{\rm Mag} = m \tilde{n} \bm{\kappa} \times  (\bm{v}_v - \bm{v}_s) = m \tilde{n} \bm{\kappa} \times \bm{v}_d,
\end{equation}
where Eq.~\eqref{eq:velocity_final} has been used to obtain the second equality. Hence, the Magnus force should only give rise to the velocity $\bm{v}_d$ resulting from the density gradient. The force balance argument used to obtain Eq.~\eqref{eq:velocity_literature_energy} is therefore called into question, since it is not clear which forces are actually being equated.

The second approach is to use a matched asymptotic expansion \cite{pismen_motion_1991, rubinstein_vortex_1994}, where analytic solutions of the Gross--Pitaevskii equation are found both within and far from the vortex core. The two solutions are then matched at an intermediate length scale, providing an analytic expression for the vortex velocity of the form \cite{svidzinsky_stability_2000}:
\begin{equation} \label{eq:velocity_literature_density}
\bm{v}_v = \frac{3\hbar}{4m\mu} \log \left( \frac{R}{\xi} \right) \hat{\bm{\kappa}} \times \nabla V_{\rm trap}.
\end{equation}
This expression can be equivalently described in terms of a density gradient \cite{sheehy_vortices_2004, nilsen_velocity_2006}, since $\nabla \tilde{n} \propto \nabla V_{\rm trap}$. Hence, this expression is mathematically equivalent to $\bm{v}_d$ in Eq.~\eqref{eq:velocity_final}, up to a correction factor. The obvious drawback of this expression is that it neglects the phase gradient velocity $\bm{v}_s$, accounting for its absence with a multiplicative factor.

\begin{figure}[tb]
\centering
\includegraphics[width=\columnwidth]{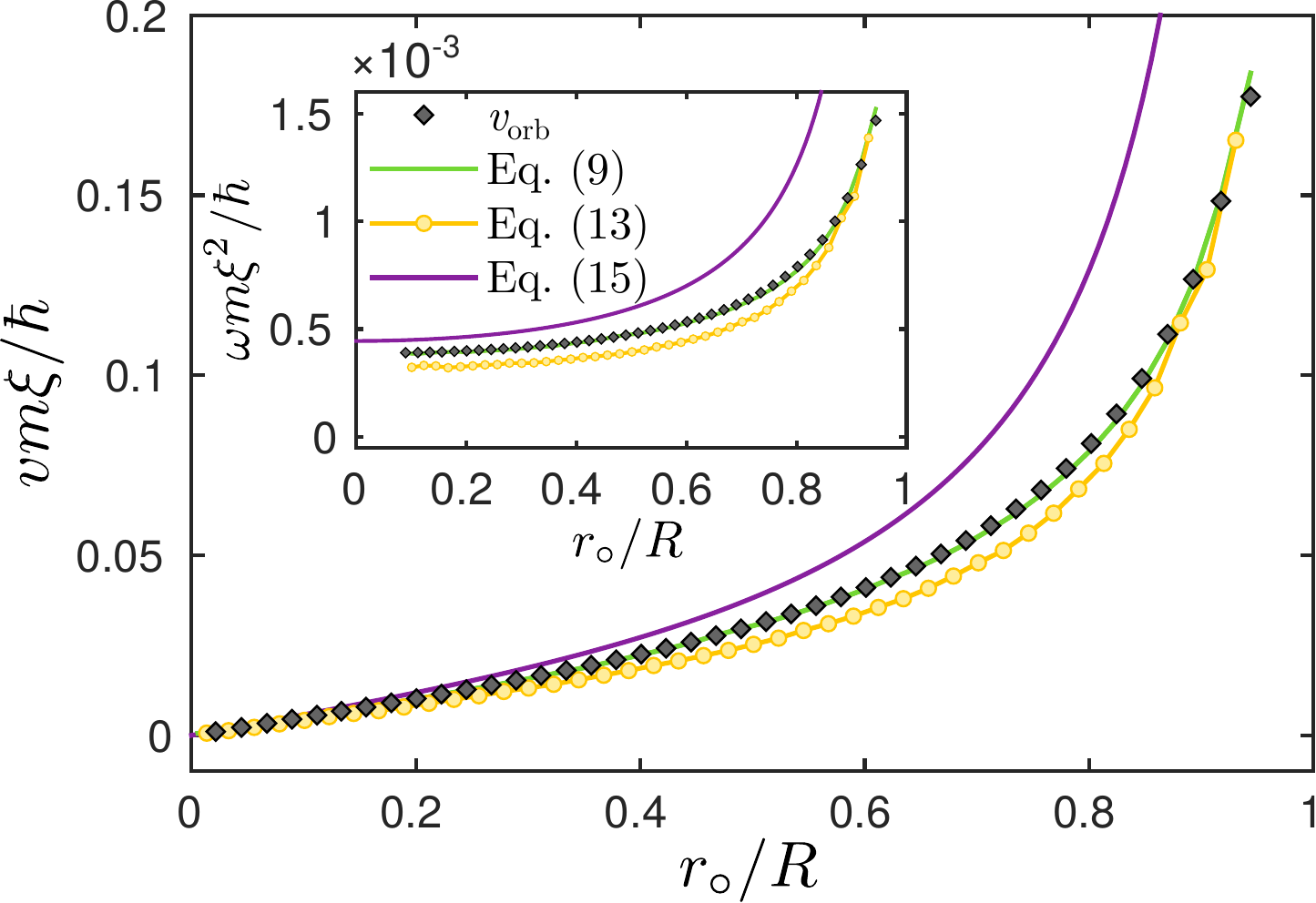}
\caption{Comparison between our numerically obtained orbital velocity (from Fig.~\ref{fig:figure_1_single}) and the predictions of Eqs.~\eqref{eq:velocity_final}, \eqref{eq:velocity_literature_energy} and \eqref{eq:velocity_literature_density} for a single vortex orbiting at radius $r_\circ$ in a harmonically trapped BEC. The inset shows the corresponding orbital frequencies, where the data for the lowest radii have been excluded due to numerical noise.}
\label{fig:figure_4_literature_comparison}
\end{figure}

For comparison between our model and those that appear in the literature, Fig.~\ref{fig:figure_4_literature_comparison} shows the orbital velocity and frequency (inset) of a vortex in a harmonic trap as calculated from Eqs.~\eqref{eq:velocity_final}, \eqref{eq:velocity_literature_energy} and \eqref{eq:velocity_literature_density} using our numerical results. Figure~\ref{fig:figure_4_literature_comparison} shows that Eq.~\eqref{eq:velocity_final} gives the best agreement with the observed orbital velocity from the GPE.

\subsection{Potential sources of confusion}
In a harmonic trap, it is possible to simplify both Eqs.~\eqref{eq:velocity_literature_energy} and \eqref{eq:velocity_literature_density} to the same functional form 
\begin{equation} \label{eq:velocity_literature_simplified}
\bm{v}_v \propto \frac{\hbar}{m} \frac{r_\circ}{R^2-r_\circ^2} \hat{\bm{\theta}}
\end{equation}
by substituting the Thomas--Fermi density profile $n(r) = n_\circ (1 - r^2/R^2)$ and local chemical potential $\mu(r) = g n(r)$, where $n_\circ$ is the density at the trap centre \cite{lundh_hydrodynamic_2000, svidzinsky_stability_2000, mcgee_rotational_2001, sheehy_vortices_2004, fetter_rotating_2009}. The agreement between these two approaches has previously been interpreted as confirmation of their validity \cite{fetter_rotating_2009}, despite the shortcomings of each method. To further confound the problem, it has also previously been assumed that Eqs.~\eqref{eq:velocity_sheehy} and \eqref{eq:velocity_literature_density} are equivalent, due to their similar functional forms \cite{sheehy_vortices_2004, fetter_rotating_2009}. However, as clarified in Sec.~\ref{sub:single_vortex_images}, these two expressions describe different physics: while Eq.~\eqref{eq:velocity_sheehy} approximates an induced phase gradient around the vortex, Eq.~\eqref{eq:velocity_literature_density} [or equivalently, the velocity $\bm{v}_d$ in Eq.~\eqref{eq:velocity_final}] describes a component of the vortex velocity that does not appear in the superfluid phase.

An additional source of potential confusion in the harmonically trapped system is that all three velocity terms in Eq.~\eqref{eq:velocity_final} have approximately the same radial dependence, as shown in Fig.~\ref{fig:figure_1_single}. Therefore, the density gradient term $\bm{v}_d$ may provide a reasonable estimate for the total velocity if multiplied by a suitable constant, as in Eq.~\eqref{eq:velocity_literature_density}. However, this approach ignores the essential physics of the induced background velocity field and image effects, and will therefore not yield quantitatively accurate results in general.

It is also worth noting that, due to the specific shape of the harmonic trapping potential, Eq.~\eqref{eq:velocity_literature_simplified} has the same functional form as predicted by the point-vortex approximation for a uniform disk of incompressible fluid; a system which corresponds to the exactly soluble electrostatic problem of a point charge inside a conducting ring. As discussed throughout Sec.~\ref{sub:results}, however, the vortex velocities in these two systems arise from different physical sources, and therefore should not be conflated.

\subsection{Image vortices}
In deriving the above expressions, Eqs.~\eqref{eq:velocity_literature_energy} and \eqref{eq:velocity_literature_density}, it is usually assumed that image vortices do not play a role in bounded inhomogeneous systems \cite{svidzinsky_stability_2000, fetter_rotating_2009}. Assuming conservation of particle number, the boundary condition for the mass current is $\hat{\textbf{n}} \cdot\textbf{j}= \hat{\textbf{n}} \cdot n \bm{v}_s=0$, where $\hat{\textbf{n}}$ is the unit vector normal to the fluid boundary. Because the density $n(\textbf{r})$ gradually approaches zero at a soft wall, this condition is automatically satisfied regardless of the value of $\bm{v}_s$ at the edge of the system. By contrast, for a hard walled system, the density is finite even at the boundary of the fluid, and therefore image vortices must be introduced to ensure $\hat{\textbf{n}} \cdot \bm{v}_{s}=0$. However, as we have argued in Sec.~\ref{sub:single_vortex_images}, there is a component of the background superfluid velocity field arising from boundary effects even in the harmonic trap, although it does not appear to be well approximated using a single localised image vortex, as is the case in the uniform disk geometry.

\subsection{Further comparisons}
Here we briefly discuss a number of other related works, whose results seem to have been largely neglected throughout the BEC literature since they were published, as most authors have instead opted to use the methods described in Sec.~\ref{sub:literature_models}.

Nilsen, Baym and Pethick \cite{nilsen_velocity_2006} obtained the same general expression for the vortex velocity in an inhomogeneous fluid, Eq.~\eqref{eq:velocity_final}, via an equivalent derivation as presented here. However, they proceeded by assuming that $\bm{v}_d=0$ and replaced $\bm{v}_s$ with $\nabla \log(\tilde{\rho})$ for a single vortex in a harmonic trap. Essentially, this lead to a model that is equivalent to Eq.~\eqref{eq:velocity_literature_density}, and which neglects important contributions to the vortex velocity.

Jezek and Cataldo \cite{jezek_vortex_2008, cataldo_influence_2009} also derived Eq.~\eqref{eq:velocity_final} using a different approach, although their model included a phenomenological correction factor multiplying $\bm{v}_d$---a factor that we have found to be unity. They also performed a detailed analysis of the induced background velocity field around a vortex in a harmonic trap \cite{jezek_vortex_2008}, as we have done in Sec.~\ref{sub:single_vortex_images}.

Various forms of Eq.~\eqref{eq:velocity_final} have also appeared in the context of optical vortex motion in nonlinear media \cite{staliunas_dynamics_1992, rozas_propagation_1997, kivshar_dynamics_1998}, since the dynamics in these optical systems are governed by a nonlinear Schr\"{o}dinger equation similar to the Gross--Pitaevskii model used here.

\section{Generalising the point-vortex model \label{sec:PVM_generalising}}

Equipped with an improved understanding of the motion of a vortex in an inhomogeneous superfluid, we now turn to an application of this theory---namely, a generalised model for describing the dynamics of point-vortices in arbitrary geometries. In particular, we will examine how our findings apply to a harmonically trapped BEC, although the approach we outline here could be applied to more general geometries. To our knowledge, all previous work considering point-vortex dynamics in harmonic traps has ignored the ambient phase gradient effects discussed throughout Secs.~\ref{sec:vortex_velocity}--\ref{sec:single_vortex_literature}. Rather, the orbital motion of a single vortex has always been modelled using the simplified form in Eq.~\eqref{eq:velocity_literature_simplified} \cite{middelkamp_stability_2010, torres_dynamics_2011, navarro_dynamics_2013}, where a multiplicative constant is included to set the timescale of the dynamics. In this Section we will show that this simplifying assumption results in a model that provides a poor quantitative description of the vortex dynamics, and that some minor adjustments based on our findings above can improve the model significantly. However, we conclude that, due to the complicated nature of the induced ambient velocity field discussed in Sec.~\ref{sub:single_vortex_images}, a fully general and efficient point-vortex description seems unachievable.

\subsection{Requirements of a point-vortex model}

We first wish to specify what we consider to be the requirements of a point-vortex model. Namely:
\begin{enumerate}[(i)]
\item The model must be simple, both computationally and conceptually. Specifically, it must be more efficient to solve numerically than the GPE, otherwise there is no improvement over the standard approach to simulating BEC dynamics. To gain the improvement, however, it may be necessary to perform initial calibrations for the model using the GPE.
\item The predictions for the velocities of each vortex in the system must only depend on their circulations and instantaneous positions.
\item The dynamics predicted by the point-vortex model must be quantitatively accurate.
\end{enumerate}

\subsection{The point-vortex model} \label{sub:PVM_derivation}

We consider a configuration of $N_v$ vortices at positions $\lbrace \textbf{r}_j(t) \rbrace$ with integer charges $\lbrace s_j \rbrace$. To obtain a point-vortex model from Eq.~\eqref{eq:velocity_final}, we need to substitute in the phase field produced by this vortex configuration, as well as the background density profile of the condensate, as a function of $\textbf{r}_j$. This approach is quite general, provided a reasonable approximation for the phase field is obtainable for the geometry under consideration. Here, we begin by demonstrating that the point-vortex model for a uniform disk can be derived exactly using Eq.~\eqref{eq:velocity_final}. We then turn to the harmonically trapped case, where an exact derivation is not possible. Instead, to arrive at a point-vortex model, we make some simplifying approximations to account for the ambient velocity fields that arise from the inhomogeneous density profile.

\subsubsection{The uniform disk system}
In the case of the uniform disk geometry, each vortex induces a single image vortex of charge $\bar{s}_j=-s_j$ located beyond the fluid boundary at position $\bar{\textbf{r}}_j=\textbf{r}_j R^2 / |\textbf{r}_j|^2$ \cite{pointin_statistical_1976, viecelli_equilibrium_1995}. Hence, the total superfluid phase is given by:
\begin{align} \label{eq:phase_field_uniform_pvm}
\phi(\textbf{r},t) = \sum_{j=1}^{N_v} \bigg\lbrace &s_j \arctan \left[ \frac{y-y_j(t)}{x-x_j(t)} \right] \nonumber \\
+ &\bar{s}_j \arctan \left[ \frac{y-\bar{y}_j(t)}{x-\bar{x}_j(t)} \right] \bigg\rbrace,
\end{align}
where the first term is produced by the physical vortices, and the second term arises from the images.
The gradient of this scalar field is:
\begin{equation}
\nabla \phi(\textbf{r},t) = \sum_{j=1}^{N_v} \left[ s_j \hat{\textbf{z}} \times \frac{(\textbf{r}-\textbf{r}_j)}{|\textbf{r}-\textbf{r}_j|^2} + \bar{s}_j \hat{\textbf{z}} \times \frac{(\textbf{r}-\bar{\textbf{r}}_j)}{|\textbf{r}-\bar{\textbf{r}}_j|^2} \right].
\end{equation}
Substituting this into Eq.~\eqref{eq:velocity_final}, and using the fact that $\nabla \log (\tilde{\rho})=0$ (due to the constant density), we find that the velocity of vortex $k$ at position $\textbf{r}_k$ is given by:
\begin{equation} \label{eq:PVM_uniform}
\bm{v}_{k} = \frac{\hbar}{m} \left[ 
\sum_{j \neq k}^{N_v} s_j \hat{\textbf{z}} \times \frac{(\textbf{r}_{k}-\textbf{r}_j)}{|\textbf{r}_{k}-\textbf{r}_j|^2}
+ \sum_{j}^{N_v} \bar{s}_j \hat{\textbf{z}} \times \frac{(\textbf{r}_{k}-\bar{\textbf{r}}_j)}{|\textbf{r}_{k}-\bar{\textbf{r}}_j|^2} \right],
\end{equation}
where the $j=k$ term in the first sum has been excluded because a vortex is not affected by its own velocity field. This is the standard point-vortex model for a disk-shaped system \cite{pointin_statistical_1976, viecelli_equilibrium_1995}: the first term describes the vortex--vortex interactions, while the second corresponds to vortex--image interactions, necessary for keeping the vortex particles within the physical boundary and ensuring that the continuity equation is satisfied there. 

\subsubsection{The harmonically trapped system}
We now move on to the more complicated case of a harmonically trapped condensate. As discussed in Sec.~\ref{sub:single_vortex_images}, the phase field induced by a vortex in an inhomogeneous condensate is nontrivial, and hence obtaining a fully general point-vortex model for this geometry is most likely not possible. Instead, our goal here is to provide improvements on the model currently used throughout the literature, without introducing significant complexity.

As shown in Fig.~\ref{fig:figure_3_vfields}(b), the ambient velocity field produced far from the vortex core for an off-centred vortex is well approximated using a standard image description (left side of the Figure). It is only in the vicinity of the vortex core that this approximation fails, as the contributions from Eq.~\eqref{eq:velocity_sheehy} become important (we ignore entirely the small effect of the multipole field discussed in Sec.~\ref{sub:multipole}). Based on this, we propose a correction to the phase field in a harmonic trap that distinguishes between self-image and non-self-image interactions. To do this, we introduce an additional set of image vortices, $\lbrace \bar{\textbf{r}}_j^\prime, \bar{s}_j^\prime \rbrace$, to produce the self-induced part of the phase field at the vortex locations $\textbf{r} = \textbf{r}_j$. In the infinitesimal region around the $k$th vortex, the phase is approximated to be:
\begin{align} \label{eq:phase_field_harmonic_pvm}
\phi_k(\textbf{r},t) = \sum_{j=1}^{N_v} s_j &\arctan \left[ \frac{y-y_j(t)}{x-x_j(t)} \right] \nonumber \\
+ \sum_{j \neq k}^{N_v} \bar{s}_j &\arctan \left[ \frac{y-\bar{y}_j(t)}{x-\bar{x}_j(t)} \right] \nonumber \\
+ \bar{s}_k^\prime &\arctan \left[ \frac{y-\bar{y}_k^\prime(t)}{x-\bar{x}_k^\prime(t)} \right],
\end{align}
while at all other locations in the fluid, the phase field is given by Eq.~\eqref{eq:phase_field_uniform_pvm}. We stress that this approach is only viable in the dilute-vortex limit when the vortices are separated well enough that the induced background velocity peak around each vortex does not significantly affect any other vortex. Alternatively, if the vortices only approach one another in relatively uniform regions of the fluid (e.g.~at the centre of the harmonic trap), the effect of Eq.~\eqref{eq:velocity_sheehy} should be negligible, and hence this approach should remain valid. To apply this double-image approximation, we substitute Eq.~\eqref{eq:phase_field_harmonic_pvm} into Eq.~\eqref{eq:velocity_final}, which yields the following point-vortex model:
\begin{align} \label{eq:PVM_harmonic_improved}
\bm{v}_{k} = \frac{\hbar}{m} \Bigg [ 
 \sum_{j \neq k}^{N_v} s_j & \hat{\textbf{z}} \times \frac{(\textbf{r}_{k}-\textbf{r}_j)}{|\textbf{r}_{k}-\textbf{r}_j|^2}
+ \sum_{j \neq k}^{N_v} \bar{s}_j \hat{\textbf{z}} \times \frac{(\textbf{r}_{k}-\bar{\textbf{r}}_j)}{|\textbf{r}_{k}-\bar{\textbf{r}}_j|^2} \nonumber \\
+ \bar{s}_k^\prime & \hat{\textbf{z}} \times \frac{(\textbf{r}_{k}-\bar{\textbf{r}}_k^\prime)}{|\textbf{r}_{k}-\bar{\textbf{r}}_k^\prime|^2}
- \hat{\bm{\kappa}} \times \nabla \log \tilde{\rho}(r_k) \Bigg ].
\end{align}
Note that we have retained the density term, since the fluid is now inhomogeneous. We approximate $\tilde{\rho}(r_k)$ using a parabolic Thomas--Fermi profile.

To obtain the generalised image description, we introduce an effective charge $\alpha$ and system radius $\sqrt{\beta} R$ for the self-images by setting $\bar{s}_j^\prime = \alpha \bar{s}_j$ and $\bar{\textbf{r}}_j^\prime = \beta \bar{\textbf{r}}_j$, respectively. For a vortex at radius $r_\circ$, this modified image will produce a velocity $v(r_\circ) =(\hbar / m) \alpha r_\circ / (\beta R^2 - r_\circ^2)$. Fitting this generalised image model to the $v_s(r_\circ)$ data in Fig.~\ref{fig:figure_1_single}, we obtain $\alpha = 6.79$, $\beta = 1.32$, which gives very good agreement with the obtained data. We therefore have all of the parameters required to test Eq.~\eqref{eq:PVM_harmonic_improved}.

\subsection{Testing the model} \label{sub:PVM_testing}

Having derived and calibrated a point-vortex model, we may test its accuracy for a few simple two-vortex scenarios to see how well it reproduces the dynamics predicted by our Gross--Pitaevskii simulations. In each scenario, we compare the performance of our model to the model used throughout the literature for a harmonically trapped BEC:
\begin{equation} \label{eq:PVM_harmonic_literature}
\bm{v}_k = \frac{\hbar}{m} \left[ 
\sum_{j \neq k}^{N_v} s_j \hat{\textbf{z}} \times \frac{(\textbf{r}_{k}-\textbf{r}_j)}{|\textbf{r}_{k}-\textbf{r}_j|^2} + \Omega_\circ \hat{\textbf{z}} \times \frac{s_k \textbf{r}_k}{R^2 - r_k^2} \right],
\end{equation}
where $\Omega_\circ = (3/2)\log(R/\xi)$ \cite{fetter_rotating_2009, middelkamp_stability_2010, navarro_dynamics_2013, middelkamp_bifurcations_2010}. The second term here corresponds to Eq.~\eqref{eq:velocity_literature_simplified}, and is responsible for the circular motion of each vortex in the system. We find that replacing $\Omega_\circ \rightarrow 0.88 \, \Omega_\circ$ gives a better prediction for the orbital frequency at the trap centre, so we use this value instead. The key differences between Eqs.~\eqref{eq:PVM_harmonic_improved} and \eqref{eq:PVM_harmonic_literature} are that (i) we include image vortex effects, and (ii) our single vortex orbital behaviour arises from the sum of the density gradient and the self-image term.

We have already examined the single vortex case in Secs.~\ref{sub:results} and \ref{sub:literature_models}. Since we have calibrated our model using the data in Fig.~\ref{fig:figure_1_single}, we find very good agreement in this case. Equation~\eqref{eq:PVM_harmonic_literature}, on the other hand, reduces to Eq.~\eqref{eq:velocity_literature_density} for a single vortex, which provides a significantly less accurate prediction, as shown in Fig.~\ref{fig:figure_4_literature_comparison}.

\begin{figure}[t]
\centering
\includegraphics[width=0.85\columnwidth]{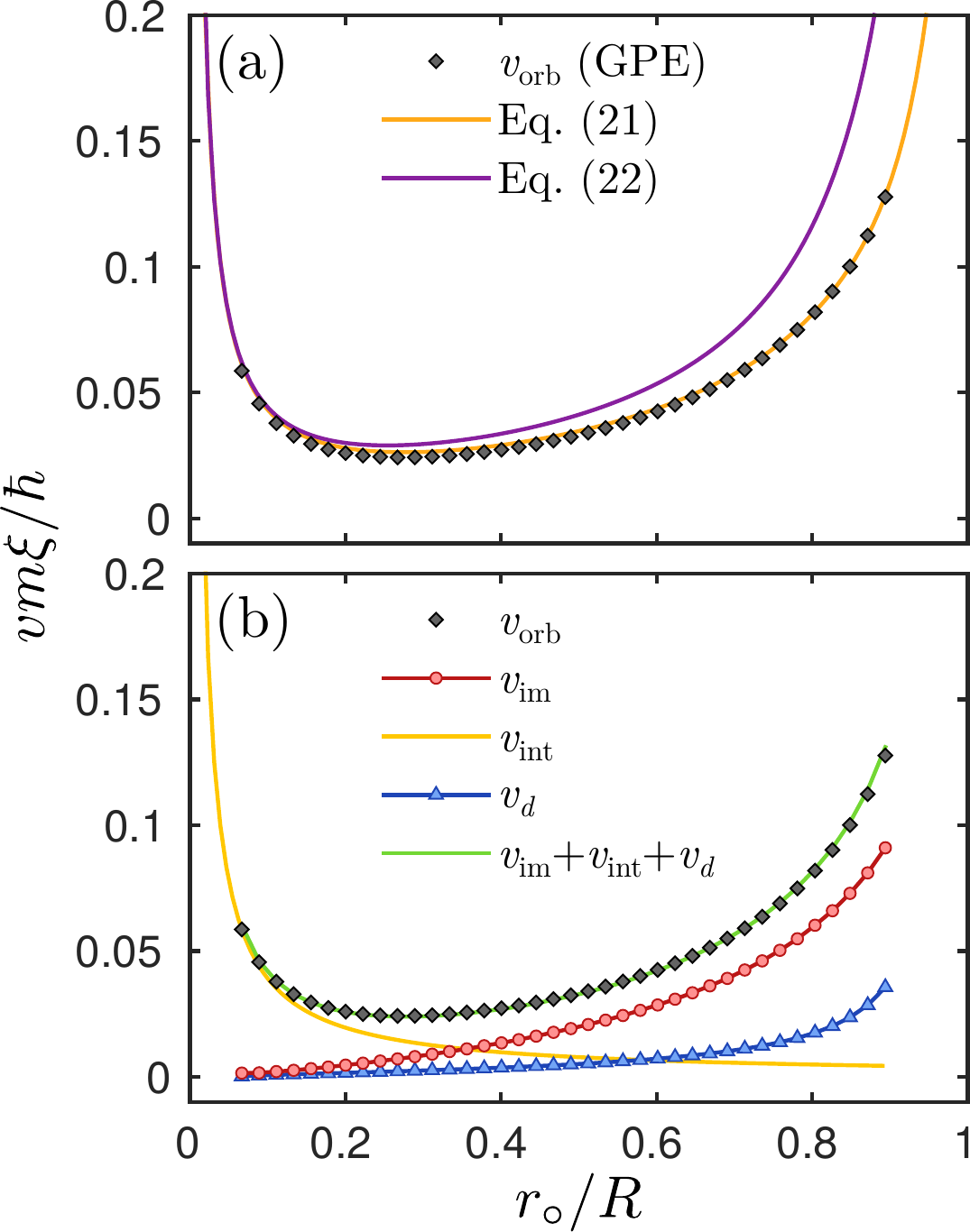}
\caption{The azimuthal velocity of two same-sign vortices in a harmonically trapped BEC as a function of their symmetric radius $r_\circ$. (a) Comparison of the orbital velocity predictions from the two point-vortex models, Eqs.~\eqref{eq:PVM_harmonic_improved} and \eqref{eq:PVM_harmonic_literature}, and the GPE. (b) Contributions to the total orbital velocity of each vortex, as measured using the GPE. We have split the ambient velocity field into $\bm{v}_s = \bm{v}_{\rm im} + \bm{v}_{\rm int}$, where $\bm{v}_{\rm im}$ is the velocity produced by the image and the density-induced phase warping, and $\bm{v}_{\rm int}$ is the velocity resulting from the vortex--vortex interaction.}
\label{fig:figure_5_two_symmetric_vortices}
\end{figure}

\subsubsection{Test I: Two symmetric same-sign vortices}

The first two-vortex case we consider is initialised with condition $s_1=s_2=1$, $\textbf{r}_1 = -\textbf{r}_2 = (x_\circ, 0)$. In this case, the two vortices symmetrically orbit around the trap centre at a constant frequency and radius. We calculate the velocity of each vortex as a function of $r_\circ$ using the GPE, and plot the separate contributions to the velocity in Fig.~\ref{fig:figure_5_two_symmetric_vortices}(b). Here, we have split the ambient velocity measurement $\bm{v}_s$ into $\bm{v}_{\rm int}(r_\circ) = 1/2r_\circ \hat{\bm{\theta}}$, the contribution from the other vortex, and $\bm{v}_{\rm im}(r_\circ)$, the velocity due to images and the density-induced phase warping. Figure \ref{fig:figure_5_two_symmetric_vortices}(a) shows how well each point-vortex model [Eqs.~\eqref{eq:PVM_harmonic_improved} and \eqref{eq:PVM_harmonic_literature}] predicts the total orbital velocity measured in the GPE. For small radii, where the vortex--vortex interaction dominates, the two predictions are equivalent; however, at larger radii our improved model is significantly more accurate.

\begin{figure}[t]
\centering
\includegraphics[width=\columnwidth]{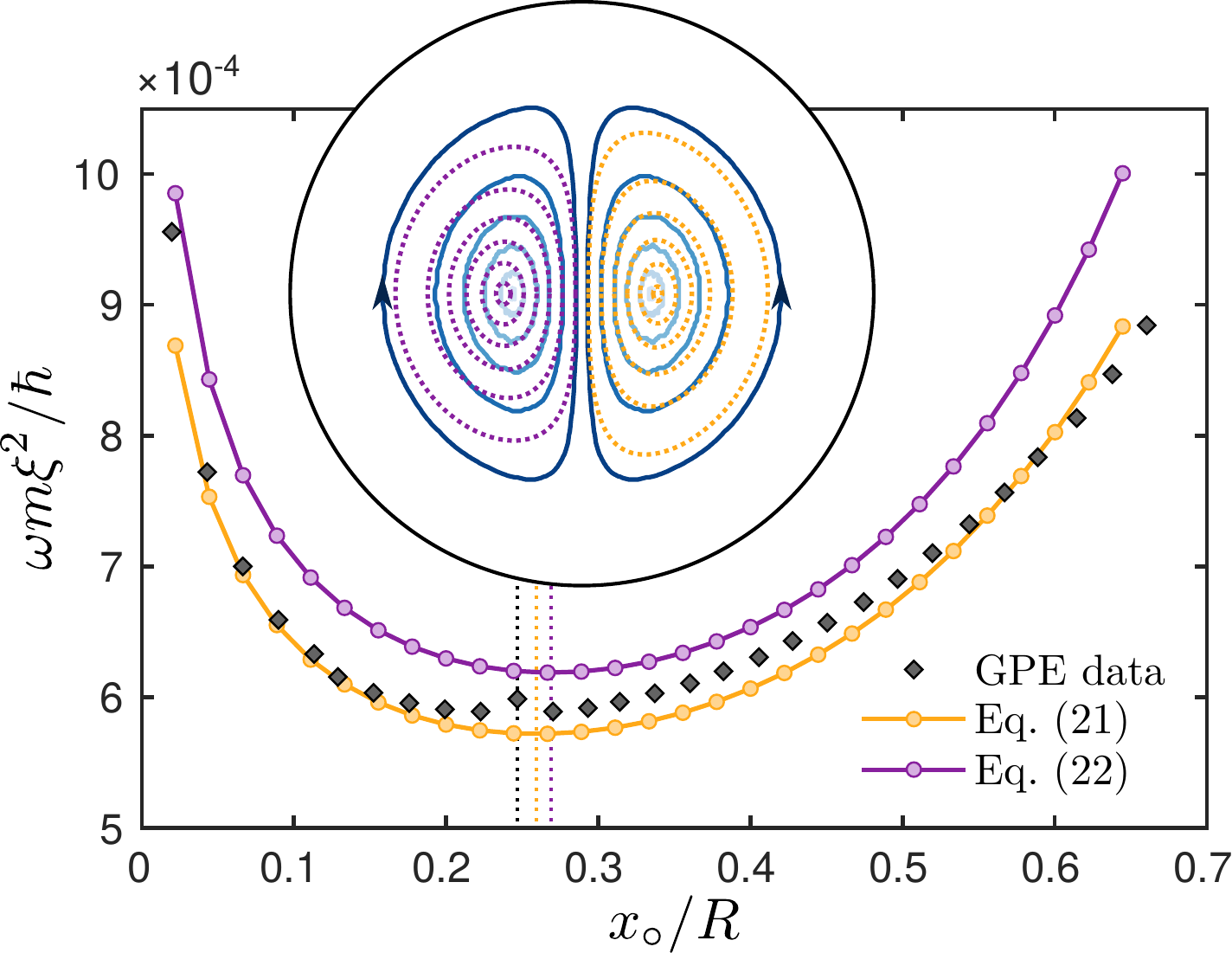}
\caption{Numerically calculated orbital frequency for a vortex dipole initiated at $\pm(x_\circ, 0)$ in a harmonically trapped BEC. The orbital frequencies calculated from the two point-vortex models, Eqs.~\eqref{eq:PVM_harmonic_improved} and \eqref{eq:PVM_harmonic_literature}, are shown alongside the Gross--Pitaevskii data for comparison. In the inset, the symmetric orbits observed in the GPE are shown as solid blue lines for the initial positions $x_\circ/R \approx \lbrace 0.02, 0.07, 0.11, 0.16, 0.20, 0.24 \rbrace$. The corresponding orbits predicted by each point-vortex model for the same initial conditions are shown as dotted lines, with Eq.~\eqref{eq:PVM_harmonic_improved} on the right, and Eq.~\eqref{eq:PVM_harmonic_literature} on the left. Dashed vertical lines in the main frame show the position of the stationary point in each model.}
\label{fig:figure_6_vortex_dipole}
\end{figure}

\subsubsection{Test II: Symmetric vortex dipole}

The second case we examine is a symmetrically placed vortex dipole, with $s_1=-s_2=1$ and initial condition $\textbf{r}_1 = -\textbf{r}_2 = (x_\circ, 0)$. For this configuration, the vortices undergo symmetric counterrotating orbits on opposite sides of the trap, which are concentric with one another as $x_\circ$ is varied. In addition, the orbits vary in frequency as a function of $x_\circ$. In Fig.~\ref{fig:figure_6_vortex_dipole}, we present both the orbits (inset) and their frequency (main frame) as a function of $x_\circ$, obtained using the GPE. For comparison, we also show the predictions from both point-vortex models, Eqs.~\eqref{eq:PVM_harmonic_improved} and \eqref{eq:PVM_harmonic_literature}. For almost all values of $x_\circ$, we obtain only a minor improvement for both the orbital shapes and their frequencies using our point-vortex model. This is not surprising, however, since this configuration violates the requirement that the vortices remain well separated while in inhomogeneous regions of the trap.

When $x_\circ \approx 0.24 \, R$, the dipole configuration is a stationary state, in which all contributions to the vortex velocity cancel. Using the two point-vortex models, Eqs.~\eqref{eq:PVM_harmonic_improved} and \eqref{eq:PVM_harmonic_literature}, this point is overestimated to be $x_\circ \approx 0.260 \, R$ and $x_\circ \approx 0.269 \, R$, respectively. Also absent from the point-vortex models is the frequency resonance observed around the stationary point in the Gross--Pitaevskii data. This resonance is the result of the compressibility not accounted for in the simplified models.

\section{Discussion \label{sec:discussion}}

We have derived a general and exact expression, Eq.~\eqref{eq:velocity_final}, for the velocity of a quantised vortex in a spatially inhomogeneous two-dimensional superfluid. Using Gross--Pitaevskii simulations, we have found that this equation provides highly accurate predictions of the velocity of vortices in some simple one- and two-vortex scenarios, both in harmonic and uniform disk-shaped traps. In doing so, we have clarified precisely how density and phase gradients affect the motion of a vortex in each of these systems. In addition, we have found a clear signature of a multipole moment induced in the velocity field of the vortex due to its internal core structure. Although past literature has made significant progress in describing vortex dynamics in nonuniform fluids, many misconceptions and erroneous assumptions exist throughout. The Magnus force has often been attributed to the total vortex velocity; however, we have shown here that it is in fact only responsible for the density gradient velocity $\bm{v}_d$ in Eq.~\eqref{eq:velocity_final}. We have also found in agreement with Ref.~\cite{jezek_vortex_2008} that image vortices, which have often been disregarded in harmonically trapped BECs, are relevant even for systems with soft boundaries.

Using our findings, we have been able to derive a new point-vortex model for a harmonically trapped BEC, which provides significant improvements for one- and two-vortex dynamics over the model currently in use throughout the literature. However, for our approach to remain quantitatively accurate, the vortices must remain dilute while in regions of varying density, since our simplified model does not rigorously account for induced ambient velocity fields in regions of varying density. Due to this stringent requirement, even with our improvements, the point-vortex model fails to provide quantitative accuracy for many simple two-vortex scenarios. Of course, the model could easily be improved by introducing more accurate approximations for the induced ambient velocity fields around each vortex; however, any added complexity may rapidly negate the simplicity required of the point-vortex model. We therefore conclude that a quantitatively accurate point-vortex treatment for arbitrary trap shapes is not possible in general due to the difficulties of modelling ambient velocity fields, which fundamentally arise from the compressibility of the fluid. For a qualitative or statistically satisfactory point-vortex model, on the other hand, the approach presented here should be straightforward to apply in a wide variety of inhomogeneous systems.

\begin{acknowledgments}
We acknowledge financial support from the Australian Postgraduate Award (A.G.), the Australian Research Council via Discovery Projects DP130102321 (T.S., K.H.) and DP170104180 (T.S.), and the nVidia research grant scheme.
\end{acknowledgments}


\begin{thebibliography}{102}%
\makeatletter
\providecommand \@ifxundefined [1]{%
 \@ifx{#1\undefined}
}%
\providecommand \@ifnum [1]{%
 \ifnum #1\expandafter \@firstoftwo
 \else \expandafter \@secondoftwo
 \fi
}%
\providecommand \@ifx [1]{%
 \ifx #1\expandafter \@firstoftwo
 \else \expandafter \@secondoftwo
 \fi
}%
\providecommand \natexlab [1]{#1}%
\providecommand \enquote  [1]{``#1''}%
\providecommand \bibnamefont  [1]{#1}%
\providecommand \bibfnamefont [1]{#1}%
\providecommand \citenamefont [1]{#1}%
\providecommand \href@noop [0]{\@secondoftwo}%
\providecommand \href [0]{\begingroup \@sanitize@url \@href}%
\providecommand \@href[1]{\@@startlink{#1}\@@href}%
\providecommand \@@href[1]{\endgroup#1\@@endlink}%
\providecommand \@sanitize@url [0]{\catcode `\\12\catcode `\$12\catcode
  `\&12\catcode `\#12\catcode `\^12\catcode `\_12\catcode `\%12\relax}%
\providecommand \@@startlink[1]{}%
\providecommand \@@endlink[0]{}%
\providecommand \url  [0]{\begingroup\@sanitize@url \@url }%
\providecommand \@url [1]{\endgroup\@href {#1}{\urlprefix }}%
\providecommand \urlprefix  [0]{URL }%
\providecommand \Eprint [0]{\href }%
\providecommand \doibase [0]{http://dx.doi.org/}%
\providecommand \selectlanguage [0]{\@gobble}%
\providecommand \bibinfo  [0]{\@secondoftwo}%
\providecommand \bibfield  [0]{\@secondoftwo}%
\providecommand \translation [1]{[#1]}%
\providecommand \BibitemOpen [0]{}%
\providecommand \bibitemStop [0]{}%
\providecommand \bibitemNoStop [0]{.\EOS\space}%
\providecommand \EOS [0]{\spacefactor3000\relax}%
\providecommand \BibitemShut  [1]{\csname bibitem#1\endcsname}%
\let\auto@bib@innerbib\@empty
\bibitem [{\citenamefont {Pismen}(1999)}]{pismen_vortices_1999}%
  \BibitemOpen
  \bibfield  {author} {\bibinfo {author} {\bibfnamefont {L.~M.}\ \bibnamefont
  {Pismen}},\ }\href@noop {} {\emph {\bibinfo {title} {Vortices in {Nonlinear}
  {Fields}: {From} {Liquid} {Crystals} to {Superfluids}, from {Non}-equilibrium
  {Patterns} to {Cosmic} {Strings}}}}\ (\bibinfo  {publisher} {Clarendon
  Press},\ \bibinfo {year} {1999})\BibitemShut {NoStop}%
\bibitem [{\citenamefont {Coullet}\ \emph {et~al.}(1989)\citenamefont
  {Coullet}, \citenamefont {Gil},\ and\ \citenamefont
  {Rocca}}]{coullet_optical_1989}%
  \BibitemOpen
  \bibfield  {author} {\bibinfo {author} {\bibfnamefont {P.}~\bibnamefont
  {Coullet}}, \bibinfo {author} {\bibfnamefont {L.}~\bibnamefont {Gil}}, \ and\
  \bibinfo {author} {\bibfnamefont {F.}~\bibnamefont {Rocca}},\ }\href
  {\doibase 10.1016/0030-4018(89)90180-6} {\bibfield  {journal} {\bibinfo
  {journal} {Opt. Commun.}\ }\textbf {\bibinfo {volume} {73}},\ \bibinfo
  {pages} {403} (\bibinfo {year} {1989})}\BibitemShut {NoStop}%
\bibitem [{\citenamefont {Dennis}\ \emph {et~al.}(2009)\citenamefont {Dennis},
  \citenamefont {O'Holleran},\ and\ \citenamefont
  {Padgett}}]{dennis_singular_2009}%
  \BibitemOpen
  \bibfield  {author} {\bibinfo {author} {\bibfnamefont {M.~R.}\ \bibnamefont
  {Dennis}}, \bibinfo {author} {\bibfnamefont {K.}~\bibnamefont {O'Holleran}},
  \ and\ \bibinfo {author} {\bibfnamefont {M.~J.}\ \bibnamefont {Padgett}},\
  }\href {\doibase 10.1016/S0079-6638(08)00205-9} {\bibfield  {journal}
  {\bibinfo  {journal} {Prog. Optics}\ }\textbf {\bibinfo {volume} {53}},\
  \bibinfo {pages} {293} (\bibinfo {year} {2009})}\BibitemShut {NoStop}%
\bibitem [{\citenamefont {Uchida}\ and\ \citenamefont
  {Tonomura}(2010)}]{uchida_generation_2010}%
  \BibitemOpen
  \bibfield  {author} {\bibinfo {author} {\bibfnamefont {M.}~\bibnamefont
  {Uchida}}\ and\ \bibinfo {author} {\bibfnamefont {A.}~\bibnamefont
  {Tonomura}},\ }\href {\doibase 10.1038/nature08904} {\bibfield  {journal}
  {\bibinfo  {journal} {Nature}\ }\textbf {\bibinfo {volume} {464}},\ \bibinfo
  {pages} {737} (\bibinfo {year} {2010})}\BibitemShut {NoStop}%
\bibitem [{\citenamefont {Verbeeck}\ \emph {et~al.}(2010)\citenamefont
  {Verbeeck}, \citenamefont {Tian},\ and\ \citenamefont
  {Schattschneider}}]{verbeeck_production_2010}%
  \BibitemOpen
  \bibfield  {author} {\bibinfo {author} {\bibfnamefont {J.}~\bibnamefont
  {Verbeeck}}, \bibinfo {author} {\bibfnamefont {H.}~\bibnamefont {Tian}}, \
  and\ \bibinfo {author} {\bibfnamefont {P.}~\bibnamefont {Schattschneider}},\
  }\href {\doibase 10.1038/nature09366} {\bibfield  {journal} {\bibinfo
  {journal} {Nature}\ }\textbf {\bibinfo {volume} {467}},\ \bibinfo {pages}
  {301} (\bibinfo {year} {2010})}\BibitemShut {NoStop}%
\bibitem [{\citenamefont {Bliokh}\ \emph {et~al.}(2017)\citenamefont {Bliokh},
  \citenamefont {Ivanov}, \citenamefont {Guzzinati}, \citenamefont {Clark},
  \citenamefont {Van Boxem}, \citenamefont {Béché}, \citenamefont
  {Juchtmans}, \citenamefont {Alonso}, \citenamefont {Schattschneider},
  \citenamefont {Nori},\ and\ \citenamefont {Verbeeck}}]{bliokh_theory_2017}%
  \BibitemOpen
  \bibfield  {author} {\bibinfo {author} {\bibfnamefont {K.~Y.}\ \bibnamefont
  {Bliokh}}, \bibinfo {author} {\bibfnamefont {I.~P.}\ \bibnamefont {Ivanov}},
  \bibinfo {author} {\bibfnamefont {G.}~\bibnamefont {Guzzinati}}, \bibinfo
  {author} {\bibfnamefont {L.}~\bibnamefont {Clark}}, \bibinfo {author}
  {\bibfnamefont {R.}~\bibnamefont {Van Boxem}}, \bibinfo {author}
  {\bibfnamefont {A.}~\bibnamefont {Béché}}, \bibinfo {author} {\bibfnamefont
  {R.}~\bibnamefont {Juchtmans}}, \bibinfo {author} {\bibfnamefont {M.~A.}\
  \bibnamefont {Alonso}}, \bibinfo {author} {\bibfnamefont {P.}~\bibnamefont
  {Schattschneider}}, \bibinfo {author} {\bibfnamefont {F.}~\bibnamefont
  {Nori}}, \ and\ \bibinfo {author} {\bibfnamefont {J.}~\bibnamefont
  {Verbeeck}},\ }\href {\doibase 10.1016/j.physrep.2017.05.006} {\bibfield
  {journal} {\bibinfo  {journal} {Phys. Rep.}\ }\textbf {\bibinfo {volume}
  {690}},\ \bibinfo {pages} {1} (\bibinfo {year} {2017})}\BibitemShut {NoStop}%
\bibitem [{\citenamefont {Blatter}\ \emph {et~al.}(1994)\citenamefont
  {Blatter}, \citenamefont {Feigel'man}, \citenamefont {Geshkenbein},
  \citenamefont {Larkin},\ and\ \citenamefont
  {Vinokur}}]{blatter_vortices_1994}%
  \BibitemOpen
  \bibfield  {author} {\bibinfo {author} {\bibfnamefont {G.}~\bibnamefont
  {Blatter}}, \bibinfo {author} {\bibfnamefont {M.~V.}\ \bibnamefont
  {Feigel'man}}, \bibinfo {author} {\bibfnamefont {V.~B.}\ \bibnamefont
  {Geshkenbein}}, \bibinfo {author} {\bibfnamefont {A.~I.}\ \bibnamefont
  {Larkin}}, \ and\ \bibinfo {author} {\bibfnamefont {V.~M.}\ \bibnamefont
  {Vinokur}},\ }\href {\doibase 10.1103/RevModPhys.66.1125} {\bibfield
  {journal} {\bibinfo  {journal} {Rev. Mod. Phys.}\ }\textbf {\bibinfo {volume}
  {66}},\ \bibinfo {pages} {1125} (\bibinfo {year} {1994})}\BibitemShut
  {NoStop}%
\bibitem [{\citenamefont {Vinen}(1961)}]{vinen_detection_1961}%
  \BibitemOpen
  \bibfield  {author} {\bibinfo {author} {\bibfnamefont {W.~F.}\ \bibnamefont
  {Vinen}},\ }\href {\doibase 10.1098/rspa.1961.0029} {\bibfield  {journal}
  {\bibinfo  {journal} {Proc. R. Soc. A}\ }\textbf {\bibinfo {volume} {260}},\
  \bibinfo {pages} {218} (\bibinfo {year} {1961})}\BibitemShut {NoStop}%
\bibitem [{\citenamefont {Donnelly}(1991)}]{donnelly_quantized_1991}%
  \BibitemOpen
  \bibfield  {author} {\bibinfo {author} {\bibfnamefont {R.~J.}\ \bibnamefont
  {Donnelly}},\ }\href@noop {} {\emph {\bibinfo {title} {Quantized vortices in
  helium {II}}}}\ (\bibinfo  {publisher} {Cambridge University Press},\
  \bibinfo {year} {1991})\BibitemShut {NoStop}%
\bibitem [{\citenamefont {Bewley}\ \emph {et~al.}(2006)\citenamefont {Bewley},
  \citenamefont {Lathrop},\ and\ \citenamefont
  {Sreenivasan}}]{bewley_superfluid_2006}%
  \BibitemOpen
  \bibfield  {author} {\bibinfo {author} {\bibfnamefont {G.~P.}\ \bibnamefont
  {Bewley}}, \bibinfo {author} {\bibfnamefont {D.~P.}\ \bibnamefont {Lathrop}},
  \ and\ \bibinfo {author} {\bibfnamefont {K.~R.}\ \bibnamefont
  {Sreenivasan}},\ }\href {\doibase 10.1038/441588a} {\bibfield  {journal}
  {\bibinfo  {journal} {Nature}\ }\textbf {\bibinfo {volume} {441}},\ \bibinfo
  {pages} {588} (\bibinfo {year} {2006})}\BibitemShut {NoStop}%
\bibitem [{\citenamefont {Nye}\ and\ \citenamefont
  {Berry}(1974)}]{nye_dislocations_1974}%
  \BibitemOpen
  \bibfield  {author} {\bibinfo {author} {\bibfnamefont {J.~F.}\ \bibnamefont
  {Nye}}\ and\ \bibinfo {author} {\bibfnamefont {M.~V.}\ \bibnamefont
  {Berry}},\ }\href {\doibase 10.1098/rspa.1974.0012} {\bibfield  {journal}
  {\bibinfo  {journal} {Proc. R. Soc. A}\ }\textbf {\bibinfo {volume} {336}},\
  \bibinfo {pages} {165} (\bibinfo {year} {1974})}\BibitemShut {NoStop}%
\bibitem [{\citenamefont {Kibble}(1976)}]{kibble_topology_1976}%
  \BibitemOpen
  \bibfield  {author} {\bibinfo {author} {\bibfnamefont {T.~W.~B.}\
  \bibnamefont {Kibble}},\ }\href {\doibase 10.1088/0305-4470/9/8/029}
  {\bibfield  {journal} {\bibinfo  {journal} {J. Phys. A: Math. Gen.}\ }\textbf
  {\bibinfo {volume} {9}},\ \bibinfo {pages} {1387} (\bibinfo {year}
  {1976})}\BibitemShut {NoStop}%
\bibitem [{\citenamefont {Zurek}(1985)}]{zurek_cosmological_1985}%
  \BibitemOpen
  \bibfield  {author} {\bibinfo {author} {\bibfnamefont {W.~H.}\ \bibnamefont
  {Zurek}},\ }\href {\doibase 10.1038/317505a0} {\bibfield  {journal} {\bibinfo
   {journal} {Nature}\ }\textbf {\bibinfo {volume} {317}},\ \bibinfo {pages}
  {505} (\bibinfo {year} {1985})}\BibitemShut {NoStop}%
\bibitem [{\citenamefont {Weiler}\ \emph {et~al.}(2008)\citenamefont {Weiler},
  \citenamefont {Neely}, \citenamefont {Scherer}, \citenamefont {Bradley},
  \citenamefont {Davis},\ and\ \citenamefont
  {Anderson}}]{weiler_spontaneous_2008}%
  \BibitemOpen
  \bibfield  {author} {\bibinfo {author} {\bibfnamefont {C.~N.}\ \bibnamefont
  {Weiler}}, \bibinfo {author} {\bibfnamefont {T.~W.}\ \bibnamefont {Neely}},
  \bibinfo {author} {\bibfnamefont {D.~R.}\ \bibnamefont {Scherer}}, \bibinfo
  {author} {\bibfnamefont {A.~S.}\ \bibnamefont {Bradley}}, \bibinfo {author}
  {\bibfnamefont {M.~J.}\ \bibnamefont {Davis}}, \ and\ \bibinfo {author}
  {\bibfnamefont {B.~P.}\ \bibnamefont {Anderson}},\ }\href {\doibase
  10.1038/nature07334} {\bibfield  {journal} {\bibinfo  {journal} {Nature}\
  }\textbf {\bibinfo {volume} {455}},\ \bibinfo {pages} {948} (\bibinfo {year}
  {2008})}\BibitemShut {NoStop}%
\bibitem [{\citenamefont {Navon}\ \emph {et~al.}(2016)\citenamefont {Navon},
  \citenamefont {Gaunt}, \citenamefont {Smith},\ and\ \citenamefont
  {Hadzibabic}}]{navon_emergence_2016}%
  \BibitemOpen
  \bibfield  {author} {\bibinfo {author} {\bibfnamefont {N.}~\bibnamefont
  {Navon}}, \bibinfo {author} {\bibfnamefont {A.~L.}\ \bibnamefont {Gaunt}},
  \bibinfo {author} {\bibfnamefont {R.~P.}\ \bibnamefont {Smith}}, \ and\
  \bibinfo {author} {\bibfnamefont {Z.}~\bibnamefont {Hadzibabic}},\ }\href
  {\doibase 10.1038/nature20114} {\bibfield  {journal} {\bibinfo  {journal}
  {Nature}\ }\textbf {\bibinfo {volume} {539}},\ \bibinfo {pages} {72}
  (\bibinfo {year} {2016})}\BibitemShut {NoStop}%
\bibitem [{\citenamefont {Matthews}\ \emph {et~al.}(1999)\citenamefont
  {Matthews}, \citenamefont {Anderson}, \citenamefont {Haljan}, \citenamefont
  {Hall}, \citenamefont {Wieman},\ and\ \citenamefont
  {Cornell}}]{matthews_vortices_1999}%
  \BibitemOpen
  \bibfield  {author} {\bibinfo {author} {\bibfnamefont {M.~R.}\ \bibnamefont
  {Matthews}}, \bibinfo {author} {\bibfnamefont {B.~P.}\ \bibnamefont
  {Anderson}}, \bibinfo {author} {\bibfnamefont {P.~C.}\ \bibnamefont
  {Haljan}}, \bibinfo {author} {\bibfnamefont {D.~S.}\ \bibnamefont {Hall}},
  \bibinfo {author} {\bibfnamefont {C.~E.}\ \bibnamefont {Wieman}}, \ and\
  \bibinfo {author} {\bibfnamefont {E.~A.}\ \bibnamefont {Cornell}},\ }\href
  {\doibase 10.1103/PhysRevLett.83.2498} {\bibfield  {journal} {\bibinfo
  {journal} {Phys. Rev. Lett.}\ }\textbf {\bibinfo {volume} {83}},\ \bibinfo
  {pages} {2498} (\bibinfo {year} {1999})}\BibitemShut {NoStop}%
\bibitem [{\citenamefont {Madison}\ \emph {et~al.}(2000)\citenamefont
  {Madison}, \citenamefont {Chevy}, \citenamefont {Wohlleben},\ and\
  \citenamefont {Dalibard}}]{madison_vortex_2000}%
  \BibitemOpen
  \bibfield  {author} {\bibinfo {author} {\bibfnamefont {K.~W.}\ \bibnamefont
  {Madison}}, \bibinfo {author} {\bibfnamefont {F.}~\bibnamefont {Chevy}},
  \bibinfo {author} {\bibfnamefont {W.}~\bibnamefont {Wohlleben}}, \ and\
  \bibinfo {author} {\bibfnamefont {J.}~\bibnamefont {Dalibard}},\ }\href
  {\doibase 10.1103/PhysRevLett.84.806} {\bibfield  {journal} {\bibinfo
  {journal} {Phys. Rev. Lett.}\ }\textbf {\bibinfo {volume} {84}},\ \bibinfo
  {pages} {806} (\bibinfo {year} {2000})}\BibitemShut {NoStop}%
\bibitem [{\citenamefont {Raman}\ \emph {et~al.}(2001)\citenamefont {Raman},
  \citenamefont {Abo-Shaeer}, \citenamefont {Vogels}, \citenamefont {Xu},\ and\
  \citenamefont {Ketterle}}]{raman_vortex_2001}%
  \BibitemOpen
  \bibfield  {author} {\bibinfo {author} {\bibfnamefont {C.}~\bibnamefont
  {Raman}}, \bibinfo {author} {\bibfnamefont {J.~R.}\ \bibnamefont
  {Abo-Shaeer}}, \bibinfo {author} {\bibfnamefont {J.~M.}\ \bibnamefont
  {Vogels}}, \bibinfo {author} {\bibfnamefont {K.}~\bibnamefont {Xu}}, \ and\
  \bibinfo {author} {\bibfnamefont {W.}~\bibnamefont {Ketterle}},\ }\href
  {\doibase 10.1103/PhysRevLett.87.210402} {\bibfield  {journal} {\bibinfo
  {journal} {Phys. Rev. Lett.}\ }\textbf {\bibinfo {volume} {87}},\ \bibinfo
  {pages} {210402} (\bibinfo {year} {2001})}\BibitemShut {NoStop}%
\bibitem [{\citenamefont {Leanhardt}\ \emph {et~al.}(2002)\citenamefont
  {Leanhardt}, \citenamefont {Görlitz}, \citenamefont {Chikkatur},
  \citenamefont {Kielpinski}, \citenamefont {Shin}, \citenamefont {Pritchard},\
  and\ \citenamefont {Ketterle}}]{leanhardt_imprinting_2002}%
  \BibitemOpen
  \bibfield  {author} {\bibinfo {author} {\bibfnamefont {A.~E.}\ \bibnamefont
  {Leanhardt}}, \bibinfo {author} {\bibfnamefont {A.}~\bibnamefont {Görlitz}},
  \bibinfo {author} {\bibfnamefont {A.~P.}\ \bibnamefont {Chikkatur}}, \bibinfo
  {author} {\bibfnamefont {D.}~\bibnamefont {Kielpinski}}, \bibinfo {author}
  {\bibfnamefont {Y.}~\bibnamefont {Shin}}, \bibinfo {author} {\bibfnamefont
  {D.~E.}\ \bibnamefont {Pritchard}}, \ and\ \bibinfo {author} {\bibfnamefont
  {W.}~\bibnamefont {Ketterle}},\ }\href {\doibase
  10.1103/PhysRevLett.89.190403} {\bibfield  {journal} {\bibinfo  {journal}
  {Phys. Rev. Lett.}\ }\textbf {\bibinfo {volume} {89}},\ \bibinfo {pages}
  {190403} (\bibinfo {year} {2002})}\BibitemShut {NoStop}%
\bibitem [{\citenamefont {Scherer}\ \emph {et~al.}(2007)\citenamefont
  {Scherer}, \citenamefont {Weiler}, \citenamefont {Neely},\ and\ \citenamefont
  {Anderson}}]{scherer_vortex_2007}%
  \BibitemOpen
  \bibfield  {author} {\bibinfo {author} {\bibfnamefont {D.~R.}\ \bibnamefont
  {Scherer}}, \bibinfo {author} {\bibfnamefont {C.~N.}\ \bibnamefont {Weiler}},
  \bibinfo {author} {\bibfnamefont {T.~W.}\ \bibnamefont {Neely}}, \ and\
  \bibinfo {author} {\bibfnamefont {B.~P.}\ \bibnamefont {Anderson}},\ }\href
  {\doibase 10.1103/PhysRevLett.98.110402} {\bibfield  {journal} {\bibinfo
  {journal} {Phys. Rev. Lett.}\ }\textbf {\bibinfo {volume} {98}},\ \bibinfo
  {pages} {110402} (\bibinfo {year} {2007})}\BibitemShut {NoStop}%
\bibitem [{\citenamefont {Wilson}\ \emph {et~al.}(2013)\citenamefont {Wilson},
  \citenamefont {Samson}, \citenamefont {Newman}, \citenamefont {Neely},\ and\
  \citenamefont {Anderson}}]{wilson_experimental_2013}%
  \BibitemOpen
  \bibfield  {author} {\bibinfo {author} {\bibfnamefont {K.~E.}\ \bibnamefont
  {Wilson}}, \bibinfo {author} {\bibfnamefont {E.~C.}\ \bibnamefont {Samson}},
  \bibinfo {author} {\bibfnamefont {Z.~L.}\ \bibnamefont {Newman}}, \bibinfo
  {author} {\bibfnamefont {T.~W.}\ \bibnamefont {Neely}}, \ and\ \bibinfo
  {author} {\bibfnamefont {B.~P.}\ \bibnamefont {Anderson}},\ }in\ \href
  {http://www.worldscientific.com/doi/abs/10.1142/9789814440400_0007} {\emph
  {\bibinfo {booktitle} {Annual {Review} of {Cold} {Atoms} and {Molecules}}}},\
  Vol.~\bibinfo {volume} {1}\ (\bibinfo  {publisher} {World Scientific},\
  \bibinfo {year} {2013})\ pp.\ \bibinfo {pages} {261--298}\BibitemShut
  {NoStop}%
\bibitem [{\citenamefont {Kwon}\ \emph {et~al.}(2016)\citenamefont {Kwon},
  \citenamefont {Kim}, \citenamefont {Seo},\ and\ \citenamefont
  {Shin}}]{kwon_observation_2016}%
  \BibitemOpen
  \bibfield  {author} {\bibinfo {author} {\bibfnamefont {W.~J.}\ \bibnamefont
  {Kwon}}, \bibinfo {author} {\bibfnamefont {J.~H.}\ \bibnamefont {Kim}},
  \bibinfo {author} {\bibfnamefont {S.~W.}\ \bibnamefont {Seo}}, \ and\
  \bibinfo {author} {\bibfnamefont {Y.}~\bibnamefont {Shin}},\ }\href {\doibase
  10.1103/PhysRevLett.117.245301} {\bibfield  {journal} {\bibinfo  {journal}
  {Phys. Rev. Lett.}\ }\textbf {\bibinfo {volume} {117}},\ \bibinfo {pages}
  {245301} (\bibinfo {year} {2016})}\BibitemShut {NoStop}%
\bibitem [{\citenamefont {Freilich}\ \emph {et~al.}(2010)\citenamefont
  {Freilich}, \citenamefont {Bianchi}, \citenamefont {Kaufman}, \citenamefont
  {Langin},\ and\ \citenamefont {Hall}}]{freilich_real-time_2010}%
  \BibitemOpen
  \bibfield  {author} {\bibinfo {author} {\bibfnamefont {D.~V.}\ \bibnamefont
  {Freilich}}, \bibinfo {author} {\bibfnamefont {D.~M.}\ \bibnamefont
  {Bianchi}}, \bibinfo {author} {\bibfnamefont {A.~M.}\ \bibnamefont
  {Kaufman}}, \bibinfo {author} {\bibfnamefont {T.~K.}\ \bibnamefont {Langin}},
  \ and\ \bibinfo {author} {\bibfnamefont {D.~S.}\ \bibnamefont {Hall}},\
  }\href {\doibase 10.1126/science.1191224} {\bibfield  {journal} {\bibinfo
  {journal} {Science}\ }\textbf {\bibinfo {volume} {329}},\ \bibinfo {pages}
  {1182} (\bibinfo {year} {2010})}\BibitemShut {NoStop}%
\bibitem [{\citenamefont {Wilson}\ \emph {et~al.}(2015)\citenamefont {Wilson},
  \citenamefont {Newman}, \citenamefont {Lowney},\ and\ \citenamefont
  {Anderson}}]{wilson_situ_2015}%
  \BibitemOpen
  \bibfield  {author} {\bibinfo {author} {\bibfnamefont {K.~E.}\ \bibnamefont
  {Wilson}}, \bibinfo {author} {\bibfnamefont {Z.~L.}\ \bibnamefont {Newman}},
  \bibinfo {author} {\bibfnamefont {J.~D.}\ \bibnamefont {Lowney}}, \ and\
  \bibinfo {author} {\bibfnamefont {B.~P.}\ \bibnamefont {Anderson}},\ }\href
  {\doibase 10.1103/PhysRevA.91.023621} {\bibfield  {journal} {\bibinfo
  {journal} {Phys. Rev. A}\ }\textbf {\bibinfo {volume} {91}},\ \bibinfo
  {pages} {023621} (\bibinfo {year} {2015})}\BibitemShut {NoStop}%
\bibitem [{\citenamefont {Seo}\ \emph {et~al.}(2017)\citenamefont {Seo},
  \citenamefont {Ko}, \citenamefont {Kim},\ and\ \citenamefont
  {Shin}}]{seo_observation_2017}%
  \BibitemOpen
  \bibfield  {author} {\bibinfo {author} {\bibfnamefont {S.~W.}\ \bibnamefont
  {Seo}}, \bibinfo {author} {\bibfnamefont {B.}~\bibnamefont {Ko}}, \bibinfo
  {author} {\bibfnamefont {J.~H.}\ \bibnamefont {Kim}}, \ and\ \bibinfo
  {author} {\bibfnamefont {Y.}~\bibnamefont {Shin}},\ }\href {\doibase
  10.1038/s41598-017-04122-9} {\bibfield  {journal} {\bibinfo  {journal} {Sci.
  Rep.}\ }\textbf {\bibinfo {volume} {7}},\ \bibinfo {pages} {4587} (\bibinfo
  {year} {2017})}\BibitemShut {NoStop}%
\bibitem [{\citenamefont {Anderson}(2010)}]{anderson_resource_2010}%
  \BibitemOpen
  \bibfield  {author} {\bibinfo {author} {\bibfnamefont {B.~P.}\ \bibnamefont
  {Anderson}},\ }\href {\doibase 10.1007/s10909-010-0224-1} {\bibfield
  {journal} {\bibinfo  {journal} {J. Low Temp. Phys.}\ }\textbf {\bibinfo
  {volume} {161}},\ \bibinfo {pages} {574} (\bibinfo {year}
  {2010})}\BibitemShut {NoStop}%
\bibitem [{\citenamefont {Anderson}\ \emph {et~al.}(2000)\citenamefont
  {Anderson}, \citenamefont {Haljan}, \citenamefont {Wieman},\ and\
  \citenamefont {Cornell}}]{anderson_vortex_2000}%
  \BibitemOpen
  \bibfield  {author} {\bibinfo {author} {\bibfnamefont {B.~P.}\ \bibnamefont
  {Anderson}}, \bibinfo {author} {\bibfnamefont {P.~C.}\ \bibnamefont
  {Haljan}}, \bibinfo {author} {\bibfnamefont {C.~E.}\ \bibnamefont {Wieman}},
  \ and\ \bibinfo {author} {\bibfnamefont {E.~A.}\ \bibnamefont {Cornell}},\
  }\href {\doibase 10.1103/PhysRevLett.85.2857} {\bibfield  {journal} {\bibinfo
   {journal} {Phys. Rev. Lett.}\ }\textbf {\bibinfo {volume} {85}},\ \bibinfo
  {pages} {2857} (\bibinfo {year} {2000})}\BibitemShut {NoStop}%
\bibitem [{\citenamefont {Bretin}\ \emph {et~al.}(2003)\citenamefont {Bretin},
  \citenamefont {Rosenbusch}, \citenamefont {Chevy}, \citenamefont
  {Shlyapnikov},\ and\ \citenamefont {Dalibard}}]{bretin_quadrupole_2003}%
  \BibitemOpen
  \bibfield  {author} {\bibinfo {author} {\bibfnamefont {V.}~\bibnamefont
  {Bretin}}, \bibinfo {author} {\bibfnamefont {P.}~\bibnamefont {Rosenbusch}},
  \bibinfo {author} {\bibfnamefont {F.}~\bibnamefont {Chevy}}, \bibinfo
  {author} {\bibfnamefont {G.~V.}\ \bibnamefont {Shlyapnikov}}, \ and\ \bibinfo
  {author} {\bibfnamefont {J.}~\bibnamefont {Dalibard}},\ }\href {\doibase
  10.1103/PhysRevLett.90.100403} {\bibfield  {journal} {\bibinfo  {journal}
  {Phys. Rev. Lett.}\ }\textbf {\bibinfo {volume} {90}},\ \bibinfo {pages}
  {100403} (\bibinfo {year} {2003})}\BibitemShut {NoStop}%
\bibitem [{\citenamefont {Hodby}\ \emph {et~al.}(2003)\citenamefont {Hodby},
  \citenamefont {Hopkins}, \citenamefont {Hechenblaikner}, \citenamefont
  {Smith},\ and\ \citenamefont {Foot}}]{hodby_experimental_2003}%
  \BibitemOpen
  \bibfield  {author} {\bibinfo {author} {\bibfnamefont {E.}~\bibnamefont
  {Hodby}}, \bibinfo {author} {\bibfnamefont {S.~A.}\ \bibnamefont {Hopkins}},
  \bibinfo {author} {\bibfnamefont {G.}~\bibnamefont {Hechenblaikner}},
  \bibinfo {author} {\bibfnamefont {N.~L.}\ \bibnamefont {Smith}}, \ and\
  \bibinfo {author} {\bibfnamefont {C.~J.}\ \bibnamefont {Foot}},\ }\href
  {\doibase 10.1103/PhysRevLett.91.090403} {\bibfield  {journal} {\bibinfo
  {journal} {Phys. Rev. Lett.}\ }\textbf {\bibinfo {volume} {91}},\ \bibinfo
  {pages} {090403} (\bibinfo {year} {2003})}\BibitemShut {NoStop}%
\bibitem [{\citenamefont {Serafini}\ \emph {et~al.}(2015)\citenamefont
  {Serafini}, \citenamefont {Barbiero}, \citenamefont {Debortoli},
  \citenamefont {Donadello}, \citenamefont {Larcher}, \citenamefont {Dalfovo},
  \citenamefont {Lamporesi},\ and\ \citenamefont
  {Ferrari}}]{serafini_dynamics_2015}%
  \BibitemOpen
  \bibfield  {author} {\bibinfo {author} {\bibfnamefont {S.}~\bibnamefont
  {Serafini}}, \bibinfo {author} {\bibfnamefont {M.}~\bibnamefont {Barbiero}},
  \bibinfo {author} {\bibfnamefont {M.}~\bibnamefont {Debortoli}}, \bibinfo
  {author} {\bibfnamefont {S.}~\bibnamefont {Donadello}}, \bibinfo {author}
  {\bibfnamefont {F.}~\bibnamefont {Larcher}}, \bibinfo {author} {\bibfnamefont
  {F.}~\bibnamefont {Dalfovo}}, \bibinfo {author} {\bibfnamefont
  {G.}~\bibnamefont {Lamporesi}}, \ and\ \bibinfo {author} {\bibfnamefont
  {G.}~\bibnamefont {Ferrari}},\ }\href {\doibase
  10.1103/PhysRevLett.115.170402} {\bibfield  {journal} {\bibinfo  {journal}
  {Phys. Rev. Lett.}\ }\textbf {\bibinfo {volume} {115}},\ \bibinfo {pages}
  {170402} (\bibinfo {year} {2015})}\BibitemShut {NoStop}%
\bibitem [{\citenamefont {Yefsah}\ \emph {et~al.}(2013)\citenamefont {Yefsah},
  \citenamefont {Sommer}, \citenamefont {Ku}, \citenamefont {Cheuk},
  \citenamefont {Ji}, \citenamefont {Bakr},\ and\ \citenamefont
  {Zwierlein}}]{yefsah_heavy_2013}%
  \BibitemOpen
  \bibfield  {author} {\bibinfo {author} {\bibfnamefont {T.}~\bibnamefont
  {Yefsah}}, \bibinfo {author} {\bibfnamefont {A.~T.}\ \bibnamefont {Sommer}},
  \bibinfo {author} {\bibfnamefont {M.~J.~H.}\ \bibnamefont {Ku}}, \bibinfo
  {author} {\bibfnamefont {L.~W.}\ \bibnamefont {Cheuk}}, \bibinfo {author}
  {\bibfnamefont {W.}~\bibnamefont {Ji}}, \bibinfo {author} {\bibfnamefont
  {W.~S.}\ \bibnamefont {Bakr}}, \ and\ \bibinfo {author} {\bibfnamefont
  {M.~W.}\ \bibnamefont {Zwierlein}},\ }\href {\doibase 10.1038/nature12338}
  {\bibfield  {journal} {\bibinfo  {journal} {Nature}\ }\textbf {\bibinfo
  {volume} {499}},\ \bibinfo {pages} {426} (\bibinfo {year}
  {2013})}\BibitemShut {NoStop}%
\bibitem [{\citenamefont {Ku}\ \emph {et~al.}(2014)\citenamefont {Ku},
  \citenamefont {Ji}, \citenamefont {Mukherjee}, \citenamefont
  {Guardado-Sanchez}, \citenamefont {Cheuk}, \citenamefont {Yefsah},\ and\
  \citenamefont {Zwierlein}}]{ku_motion_2014}%
  \BibitemOpen
  \bibfield  {author} {\bibinfo {author} {\bibfnamefont {M.~J.~H.}\
  \bibnamefont {Ku}}, \bibinfo {author} {\bibfnamefont {W.}~\bibnamefont {Ji}},
  \bibinfo {author} {\bibfnamefont {B.}~\bibnamefont {Mukherjee}}, \bibinfo
  {author} {\bibfnamefont {E.}~\bibnamefont {Guardado-Sanchez}}, \bibinfo
  {author} {\bibfnamefont {L.~W.}\ \bibnamefont {Cheuk}}, \bibinfo {author}
  {\bibfnamefont {T.}~\bibnamefont {Yefsah}}, \ and\ \bibinfo {author}
  {\bibfnamefont {M.~W.}\ \bibnamefont {Zwierlein}},\ }\href {\doibase
  10.1103/PhysRevLett.113.065301} {\bibfield  {journal} {\bibinfo  {journal}
  {Phys. Rev. Lett.}\ }\textbf {\bibinfo {volume} {113}},\ \bibinfo {pages}
  {065301} (\bibinfo {year} {2014})}\BibitemShut {NoStop}%
\bibitem [{\citenamefont {Jackson}\ \emph {et~al.}(1999)\citenamefont
  {Jackson}, \citenamefont {McCann},\ and\ \citenamefont
  {Adams}}]{jackson_vortex_1999}%
  \BibitemOpen
  \bibfield  {author} {\bibinfo {author} {\bibfnamefont {B.}~\bibnamefont
  {Jackson}}, \bibinfo {author} {\bibfnamefont {J.~F.}\ \bibnamefont {McCann}},
  \ and\ \bibinfo {author} {\bibfnamefont {C.~S.}\ \bibnamefont {Adams}},\
  }\href {\doibase 10.1103/PhysRevA.61.013604} {\bibfield  {journal} {\bibinfo
  {journal} {Phys. Rev. A}\ }\textbf {\bibinfo {volume} {61}},\ \bibinfo
  {pages} {013604} (\bibinfo {year} {1999})}\BibitemShut {NoStop}%
\bibitem [{\citenamefont {Lundh}\ and\ \citenamefont
  {Ao}(2000)}]{lundh_hydrodynamic_2000}%
  \BibitemOpen
  \bibfield  {author} {\bibinfo {author} {\bibfnamefont {E.}~\bibnamefont
  {Lundh}}\ and\ \bibinfo {author} {\bibfnamefont {P.}~\bibnamefont {Ao}},\
  }\href {\doibase 10.1103/PhysRevA.61.063612} {\bibfield  {journal} {\bibinfo
  {journal} {Phys. Rev. A}\ }\textbf {\bibinfo {volume} {61}},\ \bibinfo
  {pages} {063612} (\bibinfo {year} {2000})}\BibitemShut {NoStop}%
\bibitem [{\citenamefont {Svidzinsky}\ and\ \citenamefont
  {Fetter}(2000{\natexlab{a}})}]{svidzinsky_stability_2000}%
  \BibitemOpen
  \bibfield  {author} {\bibinfo {author} {\bibfnamefont {A.~A.}\ \bibnamefont
  {Svidzinsky}}\ and\ \bibinfo {author} {\bibfnamefont {A.~L.}\ \bibnamefont
  {Fetter}},\ }\href {\doibase 10.1103/PhysRevLett.84.5919} {\bibfield
  {journal} {\bibinfo  {journal} {Phys. Rev. Lett.}\ }\textbf {\bibinfo
  {volume} {84}},\ \bibinfo {pages} {5919} (\bibinfo {year}
  {2000}{\natexlab{a}})}\BibitemShut {NoStop}%
\bibitem [{\citenamefont {Svidzinsky}\ and\ \citenamefont
  {Fetter}(2000{\natexlab{b}})}]{svidzinsky_dynamics_2000}%
  \BibitemOpen
  \bibfield  {author} {\bibinfo {author} {\bibfnamefont {A.~A.}\ \bibnamefont
  {Svidzinsky}}\ and\ \bibinfo {author} {\bibfnamefont {A.~L.}\ \bibnamefont
  {Fetter}},\ }\href {\doibase 10.1103/PhysRevA.62.063617} {\bibfield
  {journal} {\bibinfo  {journal} {Phys. Rev. A}\ }\textbf {\bibinfo {volume}
  {62}},\ \bibinfo {pages} {063617} (\bibinfo {year}
  {2000}{\natexlab{b}})}\BibitemShut {NoStop}%
\bibitem [{\citenamefont {Fetter}\ and\ \citenamefont
  {Kim}(2001)}]{fetter_vortex_2001}%
  \BibitemOpen
  \bibfield  {author} {\bibinfo {author} {\bibfnamefont {A.~L.}\ \bibnamefont
  {Fetter}}\ and\ \bibinfo {author} {\bibfnamefont {J.-k.}\ \bibnamefont
  {Kim}},\ }\href {\doibase 10.1023/A:1012919924475} {\bibfield  {journal}
  {\bibinfo  {journal} {J. Low Temp. Phys.}\ }\textbf {\bibinfo {volume}
  {125}},\ \bibinfo {pages} {239} (\bibinfo {year} {2001})}\BibitemShut
  {NoStop}%
\bibitem [{\citenamefont {McGee}\ and\ \citenamefont
  {Holland}(2001)}]{mcgee_rotational_2001}%
  \BibitemOpen
  \bibfield  {author} {\bibinfo {author} {\bibfnamefont {S.~A.}\ \bibnamefont
  {McGee}}\ and\ \bibinfo {author} {\bibfnamefont {M.~J.}\ \bibnamefont
  {Holland}},\ }\href {\doibase 10.1103/PhysRevA.63.043608} {\bibfield
  {journal} {\bibinfo  {journal} {Phys. Rev. A}\ }\textbf {\bibinfo {volume}
  {63}},\ \bibinfo {pages} {043608} (\bibinfo {year} {2001})}\BibitemShut
  {NoStop}%
\bibitem [{\citenamefont {Anglin}(2002)}]{anglin_vortices_2002}%
  \BibitemOpen
  \bibfield  {author} {\bibinfo {author} {\bibfnamefont {J.~R.}\ \bibnamefont
  {Anglin}},\ }\href {\doibase 10.1103/PhysRevA.65.063611} {\bibfield
  {journal} {\bibinfo  {journal} {Phys. Rev. A}\ }\textbf {\bibinfo {volume}
  {65}},\ \bibinfo {pages} {063611} (\bibinfo {year} {2002})}\BibitemShut
  {NoStop}%
\bibitem [{\citenamefont {Sheehy}\ and\ \citenamefont
  {Radzihovsky}(2004)}]{sheehy_vortices_2004}%
  \BibitemOpen
  \bibfield  {author} {\bibinfo {author} {\bibfnamefont {D.~E.}\ \bibnamefont
  {Sheehy}}\ and\ \bibinfo {author} {\bibfnamefont {L.}~\bibnamefont
  {Radzihovsky}},\ }\href {\doibase 10.1103/PhysRevA.70.063620} {\bibfield
  {journal} {\bibinfo  {journal} {Phys. Rev. A}\ }\textbf {\bibinfo {volume}
  {70}},\ \bibinfo {pages} {063620} (\bibinfo {year} {2004})}\BibitemShut
  {NoStop}%
\bibitem [{\citenamefont {Al~Khawaja}(2005)}]{al_khawaja_vortex_2005}%
  \BibitemOpen
  \bibfield  {author} {\bibinfo {author} {\bibfnamefont {U.}~\bibnamefont
  {Al~Khawaja}},\ }\href {\doibase 10.1103/PhysRevA.71.063611} {\bibfield
  {journal} {\bibinfo  {journal} {Phys. Rev. A}\ }\textbf {\bibinfo {volume}
  {71}},\ \bibinfo {pages} {063611} (\bibinfo {year} {2005})}\BibitemShut
  {NoStop}%
\bibitem [{\citenamefont {Nilsen}\ \emph {et~al.}(2006)\citenamefont {Nilsen},
  \citenamefont {Baym},\ and\ \citenamefont {Pethick}}]{nilsen_velocity_2006}%
  \BibitemOpen
  \bibfield  {author} {\bibinfo {author} {\bibfnamefont {H.~M.}\ \bibnamefont
  {Nilsen}}, \bibinfo {author} {\bibfnamefont {G.}~\bibnamefont {Baym}}, \ and\
  \bibinfo {author} {\bibfnamefont {C.~J.}\ \bibnamefont {Pethick}},\ }\href
  {\doibase 10.1073/pnas.0602541103} {\bibfield  {journal} {\bibinfo  {journal}
  {PNAS}\ }\textbf {\bibinfo {volume} {103}},\ \bibinfo {pages} {7978}
  (\bibinfo {year} {2006})}\BibitemShut {NoStop}%
\bibitem [{\citenamefont {Jezek}\ and\ \citenamefont
  {Cataldo}(2008)}]{jezek_vortex_2008}%
  \BibitemOpen
  \bibfield  {author} {\bibinfo {author} {\bibfnamefont {D.~M.}\ \bibnamefont
  {Jezek}}\ and\ \bibinfo {author} {\bibfnamefont {H.~M.}\ \bibnamefont
  {Cataldo}},\ }\href {\doibase 10.1103/PhysRevA.77.043602} {\bibfield
  {journal} {\bibinfo  {journal} {Phys. Rev. A}\ }\textbf {\bibinfo {volume}
  {77}},\ \bibinfo {pages} {043602} (\bibinfo {year} {2008})}\BibitemShut
  {NoStop}%
\bibitem [{\citenamefont {Koens}\ and\ \citenamefont
  {Martin}(2012)}]{koens_perturbative_2012}%
  \BibitemOpen
  \bibfield  {author} {\bibinfo {author} {\bibfnamefont {L.}~\bibnamefont
  {Koens}}\ and\ \bibinfo {author} {\bibfnamefont {A.~M.}\ \bibnamefont
  {Martin}},\ }\href {\doibase 10.1103/PhysRevA.86.013605} {\bibfield
  {journal} {\bibinfo  {journal} {Phys. Rev. A}\ }\textbf {\bibinfo {volume}
  {86}},\ \bibinfo {pages} {013605} (\bibinfo {year} {2012})}\BibitemShut
  {NoStop}%
\bibitem [{\citenamefont {dos Santos}(2016)}]{dos_santos_hydrodynamics_2016}%
  \BibitemOpen
  \bibfield  {author} {\bibinfo {author} {\bibfnamefont {F.~E.~A.}\
  \bibnamefont {dos Santos}},\ }\href {\doibase 10.1103/PhysRevA.94.063633}
  {\bibfield  {journal} {\bibinfo  {journal} {Phys. Rev. A}\ }\textbf {\bibinfo
  {volume} {94}},\ \bibinfo {pages} {063633} (\bibinfo {year}
  {2016})}\BibitemShut {NoStop}%
\bibitem [{\citenamefont {Esposito}\ \emph {et~al.}(2017)\citenamefont
  {Esposito}, \citenamefont {Krichevsky},\ and\ \citenamefont
  {Nicolis}}]{esposito_vortex_2017}%
  \BibitemOpen
  \bibfield  {author} {\bibinfo {author} {\bibfnamefont {A.}~\bibnamefont
  {Esposito}}, \bibinfo {author} {\bibfnamefont {R.}~\bibnamefont
  {Krichevsky}}, \ and\ \bibinfo {author} {\bibfnamefont {A.}~\bibnamefont
  {Nicolis}},\ }\href {\doibase 10.1103/PhysRevA.96.033615} {\bibfield
  {journal} {\bibinfo  {journal} {Phys. Rev. A}\ }\textbf {\bibinfo {volume}
  {96}},\ \bibinfo {pages} {033615} (\bibinfo {year} {2017})}\BibitemShut
  {NoStop}%
\bibitem [{\citenamefont {Biasi}\ \emph {et~al.}(2017)\citenamefont {Biasi},
  \citenamefont {Bizoń}, \citenamefont {Craps},\ and\ \citenamefont
  {Evnin}}]{biasi_exact_2017}%
  \BibitemOpen
  \bibfield  {author} {\bibinfo {author} {\bibfnamefont {A.}~\bibnamefont
  {Biasi}}, \bibinfo {author} {\bibfnamefont {P.}~\bibnamefont {Bizoń}},
  \bibinfo {author} {\bibfnamefont {B.}~\bibnamefont {Craps}}, \ and\ \bibinfo
  {author} {\bibfnamefont {O.}~\bibnamefont {Evnin}},\ }\href {\doibase
  10.1103/PhysRevA.96.053615} {\bibfield  {journal} {\bibinfo  {journal} {Phys.
  Rev. A}\ }\textbf {\bibinfo {volume} {96}},\ \bibinfo {pages} {053615}
  (\bibinfo {year} {2017})}\BibitemShut {NoStop}%
\bibitem [{\citenamefont {Mason}\ \emph {et~al.}(2006)\citenamefont {Mason},
  \citenamefont {Berloff},\ and\ \citenamefont {Fetter}}]{mason_motion_2006}%
  \BibitemOpen
  \bibfield  {author} {\bibinfo {author} {\bibfnamefont {P.}~\bibnamefont
  {Mason}}, \bibinfo {author} {\bibfnamefont {N.~G.}\ \bibnamefont {Berloff}},
  \ and\ \bibinfo {author} {\bibfnamefont {A.~L.}\ \bibnamefont {Fetter}},\
  }\href {\doibase 10.1103/PhysRevA.74.043611} {\bibfield  {journal} {\bibinfo
  {journal} {Phys. Rev. A}\ }\textbf {\bibinfo {volume} {74}},\ \bibinfo
  {pages} {043611} (\bibinfo {year} {2006})}\BibitemShut {NoStop}%
\bibitem [{\citenamefont {Mason}\ and\ \citenamefont
  {Berloff}(2008)}]{mason_motion_2008}%
  \BibitemOpen
  \bibfield  {author} {\bibinfo {author} {\bibfnamefont {P.}~\bibnamefont
  {Mason}}\ and\ \bibinfo {author} {\bibfnamefont {N.~G.}\ \bibnamefont
  {Berloff}},\ }\href {\doibase 10.1103/PhysRevA.77.032107} {\bibfield
  {journal} {\bibinfo  {journal} {Phys. Rev. A}\ }\textbf {\bibinfo {volume}
  {77}},\ \bibinfo {pages} {032107} (\bibinfo {year} {2008})}\BibitemShut
  {NoStop}%
\bibitem [{\citenamefont {Fetter}(2009)}]{fetter_rotating_2009}%
  \BibitemOpen
  \bibfield  {author} {\bibinfo {author} {\bibfnamefont {A.~L.}\ \bibnamefont
  {Fetter}},\ }\href {\doibase 10.1103/RevModPhys.81.647} {\bibfield  {journal}
  {\bibinfo  {journal} {Rev. Mod. Phys.}\ }\textbf {\bibinfo {volume} {81}},\
  \bibinfo {pages} {647} (\bibinfo {year} {2009})}\BibitemShut {NoStop}%
\bibitem [{\citenamefont {Cataldo}\ and\ \citenamefont
  {Jezek}(2009)}]{cataldo_influence_2009}%
  \BibitemOpen
  \bibfield  {author} {\bibinfo {author} {\bibfnamefont {H.~M.}\ \bibnamefont
  {Cataldo}}\ and\ \bibinfo {author} {\bibfnamefont {D.~M.}\ \bibnamefont
  {Jezek}},\ }\href {\doibase 10.1140/epjd/e2009-00196-3} {\bibfield  {journal}
  {\bibinfo  {journal} {Eur. Phys. J. D}\ }\textbf {\bibinfo {volume} {54}},\
  \bibinfo {pages} {585} (\bibinfo {year} {2009})}\BibitemShut {NoStop}%
\bibitem [{\citenamefont {Kevrekidis}\ \emph {et~al.}(2017)\citenamefont
  {Kevrekidis}, \citenamefont {Wang}, \citenamefont {Carretero-González},
  \citenamefont {Frantzeskakis},\ and\ \citenamefont
  {Xie}}]{kevrekidis_vortex_2017}%
  \BibitemOpen
  \bibfield  {author} {\bibinfo {author} {\bibfnamefont {P.~G.}\ \bibnamefont
  {Kevrekidis}}, \bibinfo {author} {\bibfnamefont {W.}~\bibnamefont {Wang}},
  \bibinfo {author} {\bibfnamefont {R.}~\bibnamefont {Carretero-González}},
  \bibinfo {author} {\bibfnamefont {D.~J.}\ \bibnamefont {Frantzeskakis}}, \
  and\ \bibinfo {author} {\bibfnamefont {S.}~\bibnamefont {Xie}},\ }\href
  {\doibase 10.1103/PhysRevA.96.043612} {\bibfield  {journal} {\bibinfo
  {journal} {Phys. Rev. A}\ }\textbf {\bibinfo {volume} {96}},\ \bibinfo
  {pages} {043612} (\bibinfo {year} {2017})}\BibitemShut {NoStop}%
\bibitem [{\citenamefont {Henderson}\ \emph {et~al.}(2009)\citenamefont
  {Henderson}, \citenamefont {Ryu}, \citenamefont {MacCormick},\ and\
  \citenamefont {Boshier}}]{henderson_experimental_2009}%
  \BibitemOpen
  \bibfield  {author} {\bibinfo {author} {\bibfnamefont {K.}~\bibnamefont
  {Henderson}}, \bibinfo {author} {\bibfnamefont {C.}~\bibnamefont {Ryu}},
  \bibinfo {author} {\bibfnamefont {C.}~\bibnamefont {MacCormick}}, \ and\
  \bibinfo {author} {\bibfnamefont {M.~G.}\ \bibnamefont {Boshier}},\ }\href
  {\doibase 10.1088/1367-2630/11/4/043030} {\bibfield  {journal} {\bibinfo
  {journal} {New J. Phys.}\ }\textbf {\bibinfo {volume} {11}},\ \bibinfo
  {pages} {043030} (\bibinfo {year} {2009})}\BibitemShut {NoStop}%
\bibitem [{\citenamefont {Gaunt}\ \emph {et~al.}(2013)\citenamefont {Gaunt},
  \citenamefont {Schmidutz}, \citenamefont {Gotlibovych}, \citenamefont
  {Smith},\ and\ \citenamefont {Hadzibabic}}]{gaunt_bose-einstein_2013}%
  \BibitemOpen
  \bibfield  {author} {\bibinfo {author} {\bibfnamefont {A.~L.}\ \bibnamefont
  {Gaunt}}, \bibinfo {author} {\bibfnamefont {T.~F.}\ \bibnamefont
  {Schmidutz}}, \bibinfo {author} {\bibfnamefont {I.}~\bibnamefont
  {Gotlibovych}}, \bibinfo {author} {\bibfnamefont {R.~P.}\ \bibnamefont
  {Smith}}, \ and\ \bibinfo {author} {\bibfnamefont {Z.}~\bibnamefont
  {Hadzibabic}},\ }\href {\doibase 10.1103/PhysRevLett.110.200406} {\bibfield
  {journal} {\bibinfo  {journal} {Phys. Rev. Lett.}\ }\textbf {\bibinfo
  {volume} {110}},\ \bibinfo {pages} {200406} (\bibinfo {year}
  {2013})}\BibitemShut {NoStop}%
\bibitem [{\citenamefont {Gauthier}\ \emph {et~al.}(2016)\citenamefont
  {Gauthier}, \citenamefont {Lenton}, \citenamefont {Parry}, \citenamefont
  {Baker}, \citenamefont {Davis}, \citenamefont {Rubinsztein-Dunlop},\ and\
  \citenamefont {Neely}}]{gauthier_direct_2016}%
  \BibitemOpen
  \bibfield  {author} {\bibinfo {author} {\bibfnamefont {G.}~\bibnamefont
  {Gauthier}}, \bibinfo {author} {\bibfnamefont {I.}~\bibnamefont {Lenton}},
  \bibinfo {author} {\bibfnamefont {N.~M.}\ \bibnamefont {Parry}}, \bibinfo
  {author} {\bibfnamefont {M.}~\bibnamefont {Baker}}, \bibinfo {author}
  {\bibfnamefont {M.~J.}\ \bibnamefont {Davis}}, \bibinfo {author}
  {\bibfnamefont {H.}~\bibnamefont {Rubinsztein-Dunlop}}, \ and\ \bibinfo
  {author} {\bibfnamefont {T.~W.}\ \bibnamefont {Neely}},\ }\href {\doibase
  10.1364/OPTICA.3.001136} {\bibfield  {journal} {\bibinfo  {journal} {Optica}\
  }\textbf {\bibinfo {volume} {3}},\ \bibinfo {pages} {1136} (\bibinfo {year}
  {2016})}\BibitemShut {NoStop}%
\bibitem [{\citenamefont {Neely}\ \emph {et~al.}(2010)\citenamefont {Neely},
  \citenamefont {Samson}, \citenamefont {Bradley}, \citenamefont {Davis},\ and\
  \citenamefont {Anderson}}]{neely_observation_2010}%
  \BibitemOpen
  \bibfield  {author} {\bibinfo {author} {\bibfnamefont {T.~W.}\ \bibnamefont
  {Neely}}, \bibinfo {author} {\bibfnamefont {E.~C.}\ \bibnamefont {Samson}},
  \bibinfo {author} {\bibfnamefont {A.~S.}\ \bibnamefont {Bradley}}, \bibinfo
  {author} {\bibfnamefont {M.~J.}\ \bibnamefont {Davis}}, \ and\ \bibinfo
  {author} {\bibfnamefont {B.~P.}\ \bibnamefont {Anderson}},\ }\href {\doibase
  10.1103/PhysRevLett.104.160401} {\bibfield  {journal} {\bibinfo  {journal}
  {Phys. Rev. Lett.}\ }\textbf {\bibinfo {volume} {104}},\ \bibinfo {pages}
  {160401} (\bibinfo {year} {2010})}\BibitemShut {NoStop}%
\bibitem [{\citenamefont {Middelkamp}\ \emph {et~al.}(2011)\citenamefont
  {Middelkamp}, \citenamefont {Torres}, \citenamefont {Kevrekidis},
  \citenamefont {Frantzeskakis}, \citenamefont {Carretero-González},
  \citenamefont {Schmelcher}, \citenamefont {Freilich},\ and\ \citenamefont
  {Hall}}]{middelkamp_guiding-center_2011}%
  \BibitemOpen
  \bibfield  {author} {\bibinfo {author} {\bibfnamefont {S.}~\bibnamefont
  {Middelkamp}}, \bibinfo {author} {\bibfnamefont {P.~J.}\ \bibnamefont
  {Torres}}, \bibinfo {author} {\bibfnamefont {P.~G.}\ \bibnamefont
  {Kevrekidis}}, \bibinfo {author} {\bibfnamefont {D.~J.}\ \bibnamefont
  {Frantzeskakis}}, \bibinfo {author} {\bibfnamefont {R.}~\bibnamefont
  {Carretero-González}}, \bibinfo {author} {\bibfnamefont {P.}~\bibnamefont
  {Schmelcher}}, \bibinfo {author} {\bibfnamefont {D.~V.}\ \bibnamefont
  {Freilich}}, \ and\ \bibinfo {author} {\bibfnamefont {D.~S.}\ \bibnamefont
  {Hall}},\ }\href {\doibase 10.1103/PhysRevA.84.011605} {\bibfield  {journal}
  {\bibinfo  {journal} {Phys. Rev. A}\ }\textbf {\bibinfo {volume} {84}},\
  \bibinfo {pages} {011605} (\bibinfo {year} {2011})}\BibitemShut {NoStop}%
\bibitem [{\citenamefont {Seman}\ \emph {et~al.}(2010)\citenamefont {Seman},
  \citenamefont {Henn}, \citenamefont {Haque}, \citenamefont {Shiozaki},
  \citenamefont {Ramos}, \citenamefont {Caracanhas}, \citenamefont {Castilho},
  \citenamefont {Castelo~Branco}, \citenamefont {Tavares}, \citenamefont
  {Poveda-Cuevas}, \citenamefont {Roati}, \citenamefont {Magalhães},\ and\
  \citenamefont {Bagnato}}]{seman_three-vortex_2010}%
  \BibitemOpen
  \bibfield  {author} {\bibinfo {author} {\bibfnamefont {J.~A.}\ \bibnamefont
  {Seman}}, \bibinfo {author} {\bibfnamefont {E.~A.~L.}\ \bibnamefont {Henn}},
  \bibinfo {author} {\bibfnamefont {M.}~\bibnamefont {Haque}}, \bibinfo
  {author} {\bibfnamefont {R.~F.}\ \bibnamefont {Shiozaki}}, \bibinfo {author}
  {\bibfnamefont {E.~R.~F.}\ \bibnamefont {Ramos}}, \bibinfo {author}
  {\bibfnamefont {M.}~\bibnamefont {Caracanhas}}, \bibinfo {author}
  {\bibfnamefont {P.}~\bibnamefont {Castilho}}, \bibinfo {author}
  {\bibfnamefont {C.}~\bibnamefont {Castelo~Branco}}, \bibinfo {author}
  {\bibfnamefont {P.~E.~S.}\ \bibnamefont {Tavares}}, \bibinfo {author}
  {\bibfnamefont {F.~J.}\ \bibnamefont {Poveda-Cuevas}}, \bibinfo {author}
  {\bibfnamefont {G.}~\bibnamefont {Roati}}, \bibinfo {author} {\bibfnamefont
  {K.~M.~F.}\ \bibnamefont {Magalhães}}, \ and\ \bibinfo {author}
  {\bibfnamefont {V.~S.}\ \bibnamefont {Bagnato}},\ }\href {\doibase
  10.1103/PhysRevA.82.033616} {\bibfield  {journal} {\bibinfo  {journal} {Phys.
  Rev. A}\ }\textbf {\bibinfo {volume} {82}},\ \bibinfo {pages} {033616}
  (\bibinfo {year} {2010})}\BibitemShut {NoStop}%
\bibitem [{\citenamefont {Navarro}\ \emph {et~al.}(2013)\citenamefont
  {Navarro}, \citenamefont {Carretero-González}, \citenamefont {Torres},
  \citenamefont {Kevrekidis}, \citenamefont {Frantzeskakis}, \citenamefont
  {Ray}, \citenamefont {Altuntaş},\ and\ \citenamefont
  {Hall}}]{navarro_dynamics_2013}%
  \BibitemOpen
  \bibfield  {author} {\bibinfo {author} {\bibfnamefont {R.}~\bibnamefont
  {Navarro}}, \bibinfo {author} {\bibfnamefont {R.}~\bibnamefont
  {Carretero-González}}, \bibinfo {author} {\bibfnamefont {P.~J.}\
  \bibnamefont {Torres}}, \bibinfo {author} {\bibfnamefont {P.~G.}\
  \bibnamefont {Kevrekidis}}, \bibinfo {author} {\bibfnamefont {D.~J.}\
  \bibnamefont {Frantzeskakis}}, \bibinfo {author} {\bibfnamefont {M.~W.}\
  \bibnamefont {Ray}}, \bibinfo {author} {\bibfnamefont {E.}~\bibnamefont
  {Altuntaş}}, \ and\ \bibinfo {author} {\bibfnamefont {D.~S.}\ \bibnamefont
  {Hall}},\ }\href {\doibase 10.1103/PhysRevLett.110.225301} {\bibfield
  {journal} {\bibinfo  {journal} {Phys. Rev. Lett.}\ }\textbf {\bibinfo
  {volume} {110}},\ \bibinfo {pages} {225301} (\bibinfo {year}
  {2013})}\BibitemShut {NoStop}%
\bibitem [{\citenamefont {Neely}\ \emph {et~al.}(2013)\citenamefont {Neely},
  \citenamefont {Bradley}, \citenamefont {Samson}, \citenamefont {Rooney},
  \citenamefont {Wright}, \citenamefont {Law}, \citenamefont
  {Carretero-González}, \citenamefont {Kevrekidis}, \citenamefont {Davis},\
  and\ \citenamefont {Anderson}}]{neely_characteristics_2013}%
  \BibitemOpen
  \bibfield  {author} {\bibinfo {author} {\bibfnamefont {T.~W.}\ \bibnamefont
  {Neely}}, \bibinfo {author} {\bibfnamefont {A.~S.}\ \bibnamefont {Bradley}},
  \bibinfo {author} {\bibfnamefont {E.~C.}\ \bibnamefont {Samson}}, \bibinfo
  {author} {\bibfnamefont {S.~J.}\ \bibnamefont {Rooney}}, \bibinfo {author}
  {\bibfnamefont {E.~M.}\ \bibnamefont {Wright}}, \bibinfo {author}
  {\bibfnamefont {K.~J.~H.}\ \bibnamefont {Law}}, \bibinfo {author}
  {\bibfnamefont {R.}~\bibnamefont {Carretero-González}}, \bibinfo {author}
  {\bibfnamefont {P.~G.}\ \bibnamefont {Kevrekidis}}, \bibinfo {author}
  {\bibfnamefont {M.~J.}\ \bibnamefont {Davis}}, \ and\ \bibinfo {author}
  {\bibfnamefont {B.~P.}\ \bibnamefont {Anderson}},\ }\href {\doibase
  10.1103/PhysRevLett.111.235301} {\bibfield  {journal} {\bibinfo  {journal}
  {Phys. Rev. Lett.}\ }\textbf {\bibinfo {volume} {111}},\ \bibinfo {pages}
  {235301} (\bibinfo {year} {2013})}\BibitemShut {NoStop}%
\bibitem [{\citenamefont {Kwon}\ \emph {et~al.}(2014)\citenamefont {Kwon},
  \citenamefont {Moon}, \citenamefont {Choi}, \citenamefont {Seo},\ and\
  \citenamefont {Shin}}]{kwon_relaxation_2014}%
  \BibitemOpen
  \bibfield  {author} {\bibinfo {author} {\bibfnamefont {W.~J.}\ \bibnamefont
  {Kwon}}, \bibinfo {author} {\bibfnamefont {G.}~\bibnamefont {Moon}}, \bibinfo
  {author} {\bibfnamefont {J.}~\bibnamefont {Choi}}, \bibinfo {author}
  {\bibfnamefont {S.}~\bibnamefont {Seo}}, \ and\ \bibinfo {author}
  {\bibfnamefont {Y.}~\bibnamefont {Shin}},\ }\href {\doibase
  10.1103/PhysRevA.90.063627} {\bibfield  {journal} {\bibinfo  {journal} {Phys.
  Rev. A}\ }\textbf {\bibinfo {volume} {90}},\ \bibinfo {pages} {063627}
  (\bibinfo {year} {2014})}\BibitemShut {NoStop}%
\bibitem [{\citenamefont {Hess}(1967)}]{hess_angular_1967}%
  \BibitemOpen
  \bibfield  {author} {\bibinfo {author} {\bibfnamefont {G.~B.}\ \bibnamefont
  {Hess}},\ }\href {\doibase 10.1103/PhysRev.161.189} {\bibfield  {journal}
  {\bibinfo  {journal} {Phys. Rev.}\ }\textbf {\bibinfo {volume} {161}},\
  \bibinfo {pages} {189} (\bibinfo {year} {1967})}\BibitemShut {NoStop}%
\bibitem [{\citenamefont {Chang}\ \emph {et~al.}(2002)\citenamefont {Chang},
  \citenamefont {Lin},\ and\ \citenamefont {Lin}}]{chang_dynamics_2002}%
  \BibitemOpen
  \bibfield  {author} {\bibinfo {author} {\bibfnamefont {S.-M.}\ \bibnamefont
  {Chang}}, \bibinfo {author} {\bibfnamefont {W.-W.}\ \bibnamefont {Lin}}, \
  and\ \bibinfo {author} {\bibfnamefont {T.-C.}\ \bibnamefont {Lin}},\ }\href
  {\doibase 10.1142/S0218127402004644} {\bibfield  {journal} {\bibinfo
  {journal} {Int. J. Bifurc. Chaos}\ }\textbf {\bibinfo {volume} {12}},\
  \bibinfo {pages} {739} (\bibinfo {year} {2002})}\BibitemShut {NoStop}%
\bibitem [{\citenamefont {Middelkamp}\ \emph
  {et~al.}(2010{\natexlab{a}})\citenamefont {Middelkamp}, \citenamefont
  {Kevrekidis}, \citenamefont {Frantzeskakis}, \citenamefont
  {Carretero-González},\ and\ \citenamefont
  {Schmelcher}}]{middelkamp_bifurcations_2010}%
  \BibitemOpen
  \bibfield  {author} {\bibinfo {author} {\bibfnamefont {S.}~\bibnamefont
  {Middelkamp}}, \bibinfo {author} {\bibfnamefont {P.~G.}\ \bibnamefont
  {Kevrekidis}}, \bibinfo {author} {\bibfnamefont {D.~J.}\ \bibnamefont
  {Frantzeskakis}}, \bibinfo {author} {\bibfnamefont {R.}~\bibnamefont
  {Carretero-González}}, \ and\ \bibinfo {author} {\bibfnamefont
  {P.}~\bibnamefont {Schmelcher}},\ }\href {\doibase
  10.1103/PhysRevA.82.013646} {\bibfield  {journal} {\bibinfo  {journal} {Phys.
  Rev. A}\ }\textbf {\bibinfo {volume} {82}},\ \bibinfo {pages} {013646}
  (\bibinfo {year} {2010}{\natexlab{a}})}\BibitemShut {NoStop}%
\bibitem [{\citenamefont {Torres}\ \emph
  {et~al.}(2011{\natexlab{a}})\citenamefont {Torres}, \citenamefont
  {Kevrekidis}, \citenamefont {Frantzeskakis}, \citenamefont
  {Carretero-González}, \citenamefont {Schmelcher},\ and\ \citenamefont
  {Hall}}]{torres_dynamics_2011}%
  \BibitemOpen
  \bibfield  {author} {\bibinfo {author} {\bibfnamefont {P.~J.}\ \bibnamefont
  {Torres}}, \bibinfo {author} {\bibfnamefont {P.~G.}\ \bibnamefont
  {Kevrekidis}}, \bibinfo {author} {\bibfnamefont {D.~J.}\ \bibnamefont
  {Frantzeskakis}}, \bibinfo {author} {\bibfnamefont {R.}~\bibnamefont
  {Carretero-González}}, \bibinfo {author} {\bibfnamefont {P.}~\bibnamefont
  {Schmelcher}}, \ and\ \bibinfo {author} {\bibfnamefont {D.~S.}\ \bibnamefont
  {Hall}},\ }\href {\doibase 10.1016/j.physleta.2011.06.061} {\bibfield
  {journal} {\bibinfo  {journal} {Phys. Lett. A}\ }\textbf {\bibinfo {volume}
  {375}},\ \bibinfo {pages} {3044} (\bibinfo {year}
  {2011}{\natexlab{a}})}\BibitemShut {NoStop}%
\bibitem [{\citenamefont {Torres}\ \emph
  {et~al.}(2011{\natexlab{b}})\citenamefont {Torres}, \citenamefont
  {Carretero-González}, \citenamefont {Middelkamp}, \citenamefont
  {Schmelcher}, \citenamefont {Frantzeskakis},\ and\ \citenamefont
  {Kevrekidis}}]{torres_vortex_2011}%
  \BibitemOpen
  \bibfield  {author} {\bibinfo {author} {\bibfnamefont {P.~J.}\ \bibnamefont
  {Torres}}, \bibinfo {author} {\bibfnamefont {R.}~\bibnamefont
  {Carretero-González}}, \bibinfo {author} {\bibfnamefont {S.}~\bibnamefont
  {Middelkamp}}, \bibinfo {author} {\bibfnamefont {P.}~\bibnamefont
  {Schmelcher}}, \bibinfo {author} {\bibfnamefont {D.~J.}\ \bibnamefont
  {Frantzeskakis}}, \ and\ \bibinfo {author} {\bibfnamefont {P.~G.}\
  \bibnamefont {Kevrekidis}},\ }\href {http://www.ugr.es/~ptorres/docs/60.pdf}
  {\bibfield  {journal} {\bibinfo  {journal} {Comm. Pure Appl. Anal.}\ }\textbf
  {\bibinfo {volume} {10}},\ \bibinfo {pages} {1589} (\bibinfo {year}
  {2011}{\natexlab{b}})}\BibitemShut {NoStop}%
\bibitem [{\citenamefont {Simula}\ \emph {et~al.}(2014)\citenamefont {Simula},
  \citenamefont {Davis},\ and\ \citenamefont
  {Helmerson}}]{simula_emergence_2014}%
  \BibitemOpen
  \bibfield  {author} {\bibinfo {author} {\bibfnamefont {T.}~\bibnamefont
  {Simula}}, \bibinfo {author} {\bibfnamefont {M.~J.}\ \bibnamefont {Davis}}, \
  and\ \bibinfo {author} {\bibfnamefont {K.}~\bibnamefont {Helmerson}},\ }\href
  {\doibase 10.1103/PhysRevLett.113.165302} {\bibfield  {journal} {\bibinfo
  {journal} {Phys. Rev. Lett.}\ }\textbf {\bibinfo {volume} {113}},\ \bibinfo
  {pages} {165302} (\bibinfo {year} {2014})}\BibitemShut {NoStop}%
\bibitem [{\citenamefont {Murray}\ \emph {et~al.}(2016)\citenamefont {Murray},
  \citenamefont {Groszek}, \citenamefont {Kuopanportti},\ and\ \citenamefont
  {Simula}}]{murray_hamiltonian_2016}%
  \BibitemOpen
  \bibfield  {author} {\bibinfo {author} {\bibfnamefont {A.~V.}\ \bibnamefont
  {Murray}}, \bibinfo {author} {\bibfnamefont {A.~J.}\ \bibnamefont {Groszek}},
  \bibinfo {author} {\bibfnamefont {P.}~\bibnamefont {Kuopanportti}}, \ and\
  \bibinfo {author} {\bibfnamefont {T.}~\bibnamefont {Simula}},\ }\href
  {\doibase 10.1103/PhysRevA.93.033649} {\bibfield  {journal} {\bibinfo
  {journal} {Phys. Rev. A}\ }\textbf {\bibinfo {volume} {93}},\ \bibinfo
  {pages} {033649} (\bibinfo {year} {2016})}\BibitemShut {NoStop}%
\bibitem [{\citenamefont {Kim}\ \emph {et~al.}(2016)\citenamefont {Kim},
  \citenamefont {Kwon},\ and\ \citenamefont {Shin}}]{kim_role_2016}%
  \BibitemOpen
  \bibfield  {author} {\bibinfo {author} {\bibfnamefont {J.~H.}\ \bibnamefont
  {Kim}}, \bibinfo {author} {\bibfnamefont {W.~J.}\ \bibnamefont {Kwon}}, \
  and\ \bibinfo {author} {\bibfnamefont {Y.}~\bibnamefont {Shin}},\ }\href
  {\doibase 10.1103/PhysRevA.94.033612} {\bibfield  {journal} {\bibinfo
  {journal} {Phys. Rev. A}\ }\textbf {\bibinfo {volume} {94}},\ \bibinfo
  {pages} {033612} (\bibinfo {year} {2016})}\BibitemShut {NoStop}%
\bibitem [{\citenamefont {Moon}\ \emph {et~al.}(2015)\citenamefont {Moon},
  \citenamefont {Kwon}, \citenamefont {Lee},\ and\ \citenamefont
  {Shin}}]{moon_thermal_2015}%
  \BibitemOpen
  \bibfield  {author} {\bibinfo {author} {\bibfnamefont {G.}~\bibnamefont
  {Moon}}, \bibinfo {author} {\bibfnamefont {W.~J.}\ \bibnamefont {Kwon}},
  \bibinfo {author} {\bibfnamefont {H.}~\bibnamefont {Lee}}, \ and\ \bibinfo
  {author} {\bibfnamefont {Y.}~\bibnamefont {Shin}},\ }\href {\doibase
  10.1103/PhysRevA.92.051601} {\bibfield  {journal} {\bibinfo  {journal} {Phys.
  Rev. A}\ }\textbf {\bibinfo {volume} {92}},\ \bibinfo {pages} {051601}
  (\bibinfo {year} {2015})}\BibitemShut {NoStop}%
\bibitem [{\citenamefont {Billam}\ \emph {et~al.}(2015)\citenamefont {Billam},
  \citenamefont {Reeves},\ and\ \citenamefont
  {Bradley}}]{billam_spectral_2015}%
  \BibitemOpen
  \bibfield  {author} {\bibinfo {author} {\bibfnamefont {T.~P.}\ \bibnamefont
  {Billam}}, \bibinfo {author} {\bibfnamefont {M.~T.}\ \bibnamefont {Reeves}},
  \ and\ \bibinfo {author} {\bibfnamefont {A.~S.}\ \bibnamefont {Bradley}},\
  }\href {\doibase 10.1103/PhysRevA.91.023615} {\bibfield  {journal} {\bibinfo
  {journal} {Phys. Rev. A}\ }\textbf {\bibinfo {volume} {91}},\ \bibinfo
  {pages} {023615} (\bibinfo {year} {2015})}\BibitemShut {NoStop}%
\bibitem [{\citenamefont {Groszek}\ \emph {et~al.}(2018)\citenamefont
  {Groszek}, \citenamefont {Davis}, \citenamefont {Paganin}, \citenamefont
  {Helmerson},\ and\ \citenamefont {Simula}}]{groszek_vortex_2018}%
  \BibitemOpen
  \bibfield  {author} {\bibinfo {author} {\bibfnamefont {A.~J.}\ \bibnamefont
  {Groszek}}, \bibinfo {author} {\bibfnamefont {M.~J.}\ \bibnamefont {Davis}},
  \bibinfo {author} {\bibfnamefont {D.~M.}\ \bibnamefont {Paganin}}, \bibinfo
  {author} {\bibfnamefont {K.}~\bibnamefont {Helmerson}}, \ and\ \bibinfo
  {author} {\bibfnamefont {T.~P.}\ \bibnamefont {Simula}},\ }\href {\doibase
  10.1103/PhysRevLett.120.034504} {\bibfield  {journal} {\bibinfo  {journal}
  {Phys. Rev. Lett.}\ }\textbf {\bibinfo {volume} {120}},\ \bibinfo {pages}
  {034504} (\bibinfo {year} {2018})}\BibitemShut {NoStop}%
\bibitem [{\citenamefont {Šimánek}(1992)}]{simanek_adiabatic_1992}%
  \BibitemOpen
  \bibfield  {author} {\bibinfo {author} {\bibfnamefont {E.}~\bibnamefont
  {Šimánek}},\ }\href {\doibase 10.1103/PhysRevB.46.14054} {\bibfield
  {journal} {\bibinfo  {journal} {Phys. Rev. B}\ }\textbf {\bibinfo {volume}
  {46}},\ \bibinfo {pages} {14054} (\bibinfo {year} {1992})}\BibitemShut
  {NoStop}%
\bibitem [{\citenamefont {Virtanen}\ \emph
  {et~al.}(2001{\natexlab{a}})\citenamefont {Virtanen}, \citenamefont
  {Simula},\ and\ \citenamefont {Salomaa}}]{virtanen_adiabaticity_2001}%
  \BibitemOpen
  \bibfield  {author} {\bibinfo {author} {\bibfnamefont {S.~M.~M.}\
  \bibnamefont {Virtanen}}, \bibinfo {author} {\bibfnamefont {T.~P.}\
  \bibnamefont {Simula}}, \ and\ \bibinfo {author} {\bibfnamefont {M.~M.}\
  \bibnamefont {Salomaa}},\ }\href {\doibase 10.1103/PhysRevLett.87.230403}
  {\bibfield  {journal} {\bibinfo  {journal} {Phys. Rev. Lett.}\ }\textbf
  {\bibinfo {volume} {87}},\ \bibinfo {pages} {230403} (\bibinfo {year}
  {2001}{\natexlab{a}})}\BibitemShut {NoStop}%
\bibitem [{\citenamefont {Viecelli}(1995)}]{viecelli_equilibrium_1995}%
  \BibitemOpen
  \bibfield  {author} {\bibinfo {author} {\bibfnamefont {J.~A.}\ \bibnamefont
  {Viecelli}},\ }\href {\doibase 10.1063/1.868528} {\bibfield  {journal}
  {\bibinfo  {journal} {Phys. Fluids}\ }\textbf {\bibinfo {volume} {7}},\
  \bibinfo {pages} {1402} (\bibinfo {year} {1995})}\BibitemShut {NoStop}%
\bibitem [{\citenamefont {Gross}(1961)}]{gross_structure_1961}%
  \BibitemOpen
  \bibfield  {author} {\bibinfo {author} {\bibfnamefont {E.~P.}\ \bibnamefont
  {Gross}},\ }\href {\doibase 10.1007/BF02731494} {\bibfield  {journal}
  {\bibinfo  {journal} {Nuovo Cim}\ }\textbf {\bibinfo {volume} {20}},\
  \bibinfo {pages} {454} (\bibinfo {year} {1961})}\BibitemShut {NoStop}%
\bibitem [{\citenamefont {Pitaevskii}(1961)}]{pitaevskii_vortex_1961}%
  \BibitemOpen
  \bibfield  {author} {\bibinfo {author} {\bibfnamefont {L.~P.}\ \bibnamefont
  {Pitaevskii}},\ }\href {http://www.jetp.ac.ru/cgi-bin/dn/e_013_02_0451.pdf}
  {\bibfield  {journal} {\bibinfo  {journal} {J. Exp. Theor. Phys}\ }\textbf
  {\bibinfo {volume} {13}},\ \bibinfo {pages} {451} (\bibinfo {year}
  {1961})}\BibitemShut {NoStop}%
\bibitem [{\citenamefont {Pethick}\ and\ \citenamefont
  {Smith}(2008)}]{pethick_bose-einstein_2008}%
  \BibitemOpen
  \bibfield  {author} {\bibinfo {author} {\bibfnamefont {C.~J.}\ \bibnamefont
  {Pethick}}\ and\ \bibinfo {author} {\bibfnamefont {H.}~\bibnamefont
  {Smith}},\ }\href
  {http://books.google.com.au/books?hl=en&lr=&id=iBk0G3_5iIQC&oi=fnd&pg=PR11&dq=pethick+and+smith&ots=0DhGgCbONF&sig=Ok8bamPs0RiJeeTnaJ_uChoftlE}
  {\emph {\bibinfo {title} {Bose-{Einstein} condensation in dilute gases}}},\
  \bibinfo {edition} {2nd}\ ed.\ (\bibinfo  {publisher} {Cambridge University
  Press},\ \bibinfo {year} {2008})\BibitemShut {NoStop}%
\bibitem [{\citenamefont {Fetter}(1966)}]{fetter_vortices_1966}%
  \BibitemOpen
  \bibfield  {author} {\bibinfo {author} {\bibfnamefont {A.~L.}\ \bibnamefont
  {Fetter}},\ }\href {\doibase 10.1103/PhysRev.151.100} {\bibfield  {journal}
  {\bibinfo  {journal} {Phys. Rev.}\ }\textbf {\bibinfo {volume} {151}},\
  \bibinfo {pages} {100} (\bibinfo {year} {1966})}\BibitemShut {NoStop}%
\bibitem [{\citenamefont {Pointin}\ and\ \citenamefont
  {Lundgren}(1976)}]{pointin_statistical_1976}%
  \BibitemOpen
  \bibfield  {author} {\bibinfo {author} {\bibfnamefont {Y.~B.}\ \bibnamefont
  {Pointin}}\ and\ \bibinfo {author} {\bibfnamefont {T.~S.}\ \bibnamefont
  {Lundgren}},\ }\href {\doibase 10.1063/1.861347} {\bibfield  {journal}
  {\bibinfo  {journal} {Phys. Fluids}\ }\textbf {\bibinfo {volume} {19}},\
  \bibinfo {pages} {1459} (\bibinfo {year} {1976})}\BibitemShut {NoStop}%
\bibitem [{Note1()}]{Note1}%
  \BibitemOpen
  \bibinfo {note} {It was assumed in their derivation that this was the only
  contribution to the vortex orbital velocity, which we have shown is not the
  case}\BibitemShut {NoStop}%
\bibitem [{\citenamefont {Rose}(2009)}]{rose_geometrical_2009}%
  \BibitemOpen
  \bibfield  {author} {\bibinfo {author} {\bibfnamefont {H.~H.}\ \bibnamefont
  {Rose}},\ }\href {http://link.springer.com/10.1007/978-3-540-85916-1_10}
  {\emph {\bibinfo {title} {Geometrical {Charged}-{Particle} {Optics}}}}\
  (\bibinfo  {publisher} {Springer},\ \bibinfo {address} {Berlin, Heidelberg},\
  \bibinfo {year} {2009})\BibitemShut {NoStop}%
\bibitem [{\citenamefont {Dodd}\ \emph {et~al.}(1997)\citenamefont {Dodd},
  \citenamefont {Burnett}, \citenamefont {Edwards},\ and\ \citenamefont
  {Clark}}]{dodd_excitation_1997}%
  \BibitemOpen
  \bibfield  {author} {\bibinfo {author} {\bibfnamefont {R.~J.}\ \bibnamefont
  {Dodd}}, \bibinfo {author} {\bibfnamefont {K.}~\bibnamefont {Burnett}},
  \bibinfo {author} {\bibfnamefont {M.}~\bibnamefont {Edwards}}, \ and\
  \bibinfo {author} {\bibfnamefont {C.~W.}\ \bibnamefont {Clark}},\ }\href
  {\doibase 10.1103/PhysRevA.56.587} {\bibfield  {journal} {\bibinfo  {journal}
  {Phys. Rev. A}\ }\textbf {\bibinfo {volume} {56}},\ \bibinfo {pages} {587}
  (\bibinfo {year} {1997})}\BibitemShut {NoStop}%
\bibitem [{\citenamefont {Isoshima}\ and\ \citenamefont
  {Machida}(1997)}]{isoshima_bose-einstein_1997}%
  \BibitemOpen
  \bibfield  {author} {\bibinfo {author} {\bibfnamefont {T.}~\bibnamefont
  {Isoshima}}\ and\ \bibinfo {author} {\bibfnamefont {K.}~\bibnamefont
  {Machida}},\ }\href {\doibase 10.1143/JPSJ.66.3502} {\bibfield  {journal}
  {\bibinfo  {journal} {J. Phys. Soc. Jpn.}\ }\textbf {\bibinfo {volume}
  {66}},\ \bibinfo {pages} {3502} (\bibinfo {year} {1997})}\BibitemShut
  {NoStop}%
\bibitem [{\citenamefont {Virtanen}\ \emph
  {et~al.}(2001{\natexlab{b}})\citenamefont {Virtanen}, \citenamefont
  {Simula},\ and\ \citenamefont {Salomaa}}]{virtanen_structure_2001}%
  \BibitemOpen
  \bibfield  {author} {\bibinfo {author} {\bibfnamefont {S.~M.~M.}\
  \bibnamefont {Virtanen}}, \bibinfo {author} {\bibfnamefont {T.~P.}\
  \bibnamefont {Simula}}, \ and\ \bibinfo {author} {\bibfnamefont {M.~M.}\
  \bibnamefont {Salomaa}},\ }\href {\doibase 10.1103/PhysRevLett.86.2704}
  {\bibfield  {journal} {\bibinfo  {journal} {Phys. Rev. Lett.}\ }\textbf
  {\bibinfo {volume} {86}},\ \bibinfo {pages} {2704} (\bibinfo {year}
  {2001}{\natexlab{b}})}\BibitemShut {NoStop}%
\bibitem [{\citenamefont {Fetter}(2004)}]{fetter_kelvin_2004}%
  \BibitemOpen
  \bibfield  {author} {\bibinfo {author} {\bibfnamefont {A.~L.}\ \bibnamefont
  {Fetter}},\ }\href {\doibase 10.1103/PhysRevA.69.043617} {\bibfield
  {journal} {\bibinfo  {journal} {Phys. Rev. A}\ }\textbf {\bibinfo {volume}
  {69}},\ \bibinfo {pages} {043617} (\bibinfo {year} {2004})}\BibitemShut
  {NoStop}%
\bibitem [{\citenamefont {Simula}\ \emph {et~al.}(2008)\citenamefont {Simula},
  \citenamefont {Mizushima},\ and\ \citenamefont
  {Machida}}]{simula_kelvin_2008}%
  \BibitemOpen
  \bibfield  {author} {\bibinfo {author} {\bibfnamefont {T.~P.}\ \bibnamefont
  {Simula}}, \bibinfo {author} {\bibfnamefont {T.}~\bibnamefont {Mizushima}}, \
  and\ \bibinfo {author} {\bibfnamefont {K.}~\bibnamefont {Machida}},\ }\href
  {\doibase 10.1103/PhysRevLett.101.020402} {\bibfield  {journal} {\bibinfo
  {journal} {Phys. Rev. Lett.}\ }\textbf {\bibinfo {volume} {101}},\ \bibinfo
  {pages} {020402} (\bibinfo {year} {2008})}\BibitemShut {NoStop}%
\bibitem [{\citenamefont {Rooney}\ \emph {et~al.}(2011)\citenamefont {Rooney},
  \citenamefont {Blakie}, \citenamefont {Anderson},\ and\ \citenamefont
  {Bradley}}]{rooney_suppression_2011}%
  \BibitemOpen
  \bibfield  {author} {\bibinfo {author} {\bibfnamefont {S.~J.}\ \bibnamefont
  {Rooney}}, \bibinfo {author} {\bibfnamefont {P.~B.}\ \bibnamefont {Blakie}},
  \bibinfo {author} {\bibfnamefont {B.~P.}\ \bibnamefont {Anderson}}, \ and\
  \bibinfo {author} {\bibfnamefont {A.~S.}\ \bibnamefont {Bradley}},\ }\href
  {\doibase 10.1103/PhysRevA.84.023637} {\bibfield  {journal} {\bibinfo
  {journal} {Phys. Rev. A}\ }\textbf {\bibinfo {volume} {84}},\ \bibinfo
  {pages} {023637} (\bibinfo {year} {2011})}\BibitemShut {NoStop}%
\bibitem [{\citenamefont {Klein}\ \emph {et~al.}(2014)\citenamefont {Klein},
  \citenamefont {Aleiner},\ and\ \citenamefont {Agam}}]{klein_internal_2014}%
  \BibitemOpen
  \bibfield  {author} {\bibinfo {author} {\bibfnamefont {A.}~\bibnamefont
  {Klein}}, \bibinfo {author} {\bibfnamefont {I.~L.}\ \bibnamefont {Aleiner}},
  \ and\ \bibinfo {author} {\bibfnamefont {O.}~\bibnamefont {Agam}},\ }\href
  {\doibase 10.1016/j.aop.2014.04.018} {\bibfield  {journal} {\bibinfo
  {journal} {Ann. Phys.}\ }\textbf {\bibinfo {volume} {346}},\ \bibinfo {pages}
  {195} (\bibinfo {year} {2014})}\BibitemShut {NoStop}%
\bibitem [{Note2()}]{Note2}%
  \BibitemOpen
  \bibinfo {note} {Strictly, the induced multipole moments are intrinsic to the
  vortex `particle' and could therefore be removed from the phase field before
  calculating the smooth background field $\protect \bm {v}_s = \nabla \protect
  \mathaccentV {tilde}07E{\phi }$ which drives the vortex motion. However,
  since we have only subtracted the circularly symmetric monopole component
  $\protect \bm {v}_i^{(1)}(\protect \textbf {r})$, the higher order multipole
  contributions remain in our measured `background' field $\protect \bm
  {v}_s(\protect \textbf {r})$.}\BibitemShut {Stop}%
\bibitem [{\citenamefont {Rokhsar}(1997)}]{rokhsar_vortex_1997}%
  \BibitemOpen
  \bibfield  {author} {\bibinfo {author} {\bibfnamefont {D.~S.}\ \bibnamefont
  {Rokhsar}},\ }\href {\doibase 10.1103/PhysRevLett.79.2164} {\bibfield
  {journal} {\bibinfo  {journal} {Phys. Rev. Lett.}\ }\textbf {\bibinfo
  {volume} {79}},\ \bibinfo {pages} {2164} (\bibinfo {year}
  {1997})}\BibitemShut {NoStop}%
\bibitem [{\citenamefont {Toikka}\ and\ \citenamefont
  {Brand}(2017)}]{toikka_asymptotically_2017}%
  \BibitemOpen
  \bibfield  {author} {\bibinfo {author} {\bibfnamefont {L.~A.}\ \bibnamefont
  {Toikka}}\ and\ \bibinfo {author} {\bibfnamefont {J.}~\bibnamefont {Brand}},\
  }\href {\doibase 10.1088/1367-2630/aa5668} {\bibfield  {journal} {\bibinfo
  {journal} {New J. Phys.}\ }\textbf {\bibinfo {volume} {19}},\ \bibinfo
  {pages} {023029} (\bibinfo {year} {2017})}\BibitemShut {NoStop}%
\bibitem [{\citenamefont {Guenther}\ \emph {et~al.}(2017)\citenamefont
  {Guenther}, \citenamefont {Massignan},\ and\ \citenamefont
  {Fetter}}]{guenther_quantized_2017}%
  \BibitemOpen
  \bibfield  {author} {\bibinfo {author} {\bibfnamefont {N.-E.}\ \bibnamefont
  {Guenther}}, \bibinfo {author} {\bibfnamefont {P.}~\bibnamefont {Massignan}},
  \ and\ \bibinfo {author} {\bibfnamefont {A.~L.}\ \bibnamefont {Fetter}},\
  }\href {\doibase 10.1103/PhysRevA.96.063608} {\bibfield  {journal} {\bibinfo
  {journal} {Phys. Rev. A}\ }\textbf {\bibinfo {volume} {96}},\ \bibinfo
  {pages} {063608} (\bibinfo {year} {2017})}\BibitemShut {NoStop}%
\bibitem [{\citenamefont {Ambegaokar}\ \emph {et~al.}(1980)\citenamefont
  {Ambegaokar}, \citenamefont {Halperin}, \citenamefont {Nelson},\ and\
  \citenamefont {Siggia}}]{ambegaokar_dynamics_1980}%
  \BibitemOpen
  \bibfield  {author} {\bibinfo {author} {\bibfnamefont {V.}~\bibnamefont
  {Ambegaokar}}, \bibinfo {author} {\bibfnamefont {B.~I.}\ \bibnamefont
  {Halperin}}, \bibinfo {author} {\bibfnamefont {D.~R.}\ \bibnamefont
  {Nelson}}, \ and\ \bibinfo {author} {\bibfnamefont {E.~D.}\ \bibnamefont
  {Siggia}},\ }\href {\doibase 10.1103/PhysRevB.21.1806} {\bibfield  {journal}
  {\bibinfo  {journal} {Phys. Rev. B}\ }\textbf {\bibinfo {volume} {21}},\
  \bibinfo {pages} {1806} (\bibinfo {year} {1980})}\BibitemShut {NoStop}%
\bibitem [{\citenamefont {Ao}\ and\ \citenamefont
  {Thouless}(1993)}]{ao_berrys_1993}%
  \BibitemOpen
  \bibfield  {author} {\bibinfo {author} {\bibfnamefont {P.}~\bibnamefont
  {Ao}}\ and\ \bibinfo {author} {\bibfnamefont {D.~J.}\ \bibnamefont
  {Thouless}},\ }\href {\doibase 10.1103/PhysRevLett.70.2158} {\bibfield
  {journal} {\bibinfo  {journal} {Phys. Rev. Lett.}\ }\textbf {\bibinfo
  {volume} {70}},\ \bibinfo {pages} {2158} (\bibinfo {year}
  {1993})}\BibitemShut {NoStop}%
\bibitem [{\citenamefont {Thouless}\ \emph {et~al.}(1996)\citenamefont
  {Thouless}, \citenamefont {Ao},\ and\ \citenamefont
  {Niu}}]{thouless_transverse_1996}%
  \BibitemOpen
  \bibfield  {author} {\bibinfo {author} {\bibfnamefont {D.~J.}\ \bibnamefont
  {Thouless}}, \bibinfo {author} {\bibfnamefont {P.}~\bibnamefont {Ao}}, \ and\
  \bibinfo {author} {\bibfnamefont {Q.}~\bibnamefont {Niu}},\ }\href {\doibase
  10.1103/PhysRevLett.76.3758} {\bibfield  {journal} {\bibinfo  {journal}
  {Phys. Rev. Lett.}\ }\textbf {\bibinfo {volume} {76}},\ \bibinfo {pages}
  {3758} (\bibinfo {year} {1996})}\BibitemShut {NoStop}%
\bibitem [{\citenamefont {Pismen}\ and\ \citenamefont
  {Rubinstein}(1991)}]{pismen_motion_1991}%
  \BibitemOpen
  \bibfield  {author} {\bibinfo {author} {\bibfnamefont {L.~M.}\ \bibnamefont
  {Pismen}}\ and\ \bibinfo {author} {\bibfnamefont {J.}~\bibnamefont
  {Rubinstein}},\ }\href {\doibase 10.1016/0167-2789(91)90035-8} {\bibfield
  {journal} {\bibinfo  {journal} {Physica D}\ }\textbf {\bibinfo {volume}
  {47}},\ \bibinfo {pages} {353} (\bibinfo {year} {1991})}\BibitemShut
  {NoStop}%
\bibitem [{\citenamefont {Rubinstein}\ and\ \citenamefont
  {Pismen}(1994)}]{rubinstein_vortex_1994}%
  \BibitemOpen
  \bibfield  {author} {\bibinfo {author} {\bibfnamefont {B.~Y.}\ \bibnamefont
  {Rubinstein}}\ and\ \bibinfo {author} {\bibfnamefont {L.~M.}\ \bibnamefont
  {Pismen}},\ }\href {\doibase 10.1016/0167-2789(94)00119-7} {\bibfield
  {journal} {\bibinfo  {journal} {Physica D}\ }\textbf {\bibinfo {volume}
  {78}},\ \bibinfo {pages} {1} (\bibinfo {year} {1994})}\BibitemShut {NoStop}%
\bibitem [{\citenamefont {Staliunas}(1992)}]{staliunas_dynamics_1992}%
  \BibitemOpen
  \bibfield  {author} {\bibinfo {author} {\bibfnamefont {K.}~\bibnamefont
  {Staliunas}},\ }\href {\doibase 10.1016/0030-4018(92)90342-O} {\bibfield
  {journal} {\bibinfo  {journal} {Opt. Commun.}\ }\textbf {\bibinfo {volume}
  {90}},\ \bibinfo {pages} {123} (\bibinfo {year} {1992})}\BibitemShut
  {NoStop}%
\bibitem [{\citenamefont {Rozas}\ \emph {et~al.}(1997)\citenamefont {Rozas},
  \citenamefont {Law},\ and\ \citenamefont
  {Swartzlander}}]{rozas_propagation_1997}%
  \BibitemOpen
  \bibfield  {author} {\bibinfo {author} {\bibfnamefont {D.}~\bibnamefont
  {Rozas}}, \bibinfo {author} {\bibfnamefont {C.~T.}\ \bibnamefont {Law}}, \
  and\ \bibinfo {author} {\bibfnamefont {G.~A.}\ \bibnamefont {Swartzlander}},\
  }\href {\doibase 10.1364/JOSAB.14.003054} {\bibfield  {journal} {\bibinfo
  {journal} {J. Opt. Soc. Am. B}\ }\textbf {\bibinfo {volume} {14}},\ \bibinfo
  {pages} {3054} (\bibinfo {year} {1997})}\BibitemShut {NoStop}%
\bibitem [{\citenamefont {Kivshar}\ \emph {et~al.}(1998)\citenamefont
  {Kivshar}, \citenamefont {Christou}, \citenamefont {Tikhonenko},
  \citenamefont {Luther-Davies},\ and\ \citenamefont
  {Pismen}}]{kivshar_dynamics_1998}%
  \BibitemOpen
  \bibfield  {author} {\bibinfo {author} {\bibfnamefont {Y.~S.}\ \bibnamefont
  {Kivshar}}, \bibinfo {author} {\bibfnamefont {J.}~\bibnamefont {Christou}},
  \bibinfo {author} {\bibfnamefont {V.}~\bibnamefont {Tikhonenko}}, \bibinfo
  {author} {\bibfnamefont {B.}~\bibnamefont {Luther-Davies}}, \ and\ \bibinfo
  {author} {\bibfnamefont {L.~M.}\ \bibnamefont {Pismen}},\ }\href {\doibase
  10.1016/S0030-4018(98)00149-7} {\bibfield  {journal} {\bibinfo  {journal}
  {Opt. Commun.}\ }\textbf {\bibinfo {volume} {152}},\ \bibinfo {pages} {198}
  (\bibinfo {year} {1998})}\BibitemShut {NoStop}%
\bibitem [{\citenamefont {Middelkamp}\ \emph
  {et~al.}(2010{\natexlab{b}})\citenamefont {Middelkamp}, \citenamefont
  {Kevrekidis}, \citenamefont {Frantzeskakis}, \citenamefont
  {Carretero-González},\ and\ \citenamefont
  {Schmelcher}}]{middelkamp_stability_2010}%
  \BibitemOpen
  \bibfield  {author} {\bibinfo {author} {\bibfnamefont {S.}~\bibnamefont
  {Middelkamp}}, \bibinfo {author} {\bibfnamefont {P.~G.}\ \bibnamefont
  {Kevrekidis}}, \bibinfo {author} {\bibfnamefont {D.~J.}\ \bibnamefont
  {Frantzeskakis}}, \bibinfo {author} {\bibfnamefont {R.}~\bibnamefont
  {Carretero-González}}, \ and\ \bibinfo {author} {\bibfnamefont
  {P.}~\bibnamefont {Schmelcher}},\ }\href {\doibase
  10.1088/0953-4075/43/15/155303} {\bibfield  {journal} {\bibinfo  {journal}
  {J. Phys. B: At. Mol. Opt. Phys.}\ }\textbf {\bibinfo {volume} {43}},\
  \bibinfo {pages} {155303} (\bibinfo {year} {2010}{\natexlab{b}})}\BibitemShut
  {NoStop}%
\end{thebibliography}

%

\end{document}